\documentclass[aastex,superscriptaddress,showpacs,letterpaper,showkeys,preprintnumbers,altaffilletter,amssymb,amsmath,amsfonts,aps]{emulateapj}

\usepackage{graphicx}
\usepackage{amsmath}

\usepackage{natbib}
\def\be{\begin{equation}}
\def\ee{\end{equation}}
\def\ben{$$}
\def\een{$$}
\def\bea{\begin{eqnarray}}
\def\eea{\end{eqnarray}}
\def\bean{\begin{eqnarray*}}
\def\eean{\end{eqnarray*}}

\def\bi{\begin{itemize}}
\def\ei{\end{itemize}}
\def\ben{\begin{enumerate}}
\def\een{\end{enumerate}}
\def\Ep{{E_+}}
\def\Ip{{I_+}}
\def\Ec{{E_\times}}
\def\Ic{{I_\times}}

\def\En{{E_\mathrm{null}}}
\def\In{{I_\mathrm{null}}}

\def\hp{{h_+}}
\def\hc{{h_\times}}
\def\xp{{\textsc{X-Pipeline}}}
\def\matlab{{\textsc{matlab}}}
\def\data{\boldsymbol{d}}

\def\Fplus{\boldsymbol{f}^{+,\mathrm{DPF}}}
\def\Fcross{\boldsymbol{f}^{\times,\mathrm{DPF}}}

\shorttitle{Search for GWBs associated with GRBs using LIGO and Virgo}
\shortauthors{Abbott et al.}

\begin{document}

\title{Search for gravitational-wave bursts associated with gamma-ray bursts using data from LIGO Science Run 5 and Virgo Science Run 1}

\author{B.~P.~Abbott$^{28}$, 
R.~Abbott$^{28}$, 
F.~Acernese$^{18ac}$, 
R.~Adhikari$^{28}$, 
P.~Ajith$^{2}$, 
B.~Allen$^{2,75}$, 
G.~Allen$^{51}$, 
M.~Alshourbagy$^{20ab}$, 
R.~S.~Amin$^{33}$, 
S.~B.~Anderson$^{28}$
% ...
}
% ...
\address{$^{2}$Albert-Einstein-Institut, Max-Planck-Institut f\"ur Gravitationsphysik, D-30167 Hannover, Germany}
% ...
\address{$^{18}$INFN, sezione di Napoli $^a$; Universit\`a di Napoli 'Federico II'$^b$ Complesso Universitario di Monte S.Angelo, I-80126 Napoli; Universit\`a di Salerno, Fisciano, I-84084 Salerno$^c$, Italy}
% ...
\address{$^{20}$INFN, Sezione di Pisa$^a$; Universit\`a di Pisa$^b$; I-56127 Pisa; Universit\`a di Siena, I-53100 Siena$^c$, Italy}
% ...
\address{$^{28}$LIGO - California Institute of Technology, Pasadena, CA  91125, USA }
% ...
\address{$^{33}$Louisiana State University, Baton Rouge, LA  70803, USA }
% ...
\address{$^{51}$Stanford University, Stanford, CA  94305, USA }
% ...
\address{$^{75}$University of Wisconsin-Milwaukee, Milwaukee, WI  53201, USA }
% ...

\begin{abstract}

We present the results of a search for gravitational-wave bursts associated 
with 137 gamma-ray bursts (GRBs) that were detected by satellite-based 
gamma-ray experiments during the fifth LIGO science run and first Virgo 
science run.  The data used in this analysis were collected from 
2005 November 4 to 2007 October 1, and most of the GRB triggers were from 
the {\em Swift} satellite.  The search uses a coherent network analysis method 
that takes into account the different locations and orientations of the 
interferometers at the three LIGO-Virgo sites.  We find no evidence for 
gravitational-wave burst signals associated with this sample of GRBs.  
Using simulated short-duration ($<1$ s) waveforms, we set upper limits on the 
amplitude of gravitational waves associated with each GRB.  We also place 
lower bounds on the distance to each GRB under the assumption of 
a fixed energy emission in gravitational waves, with a median limit of 
\mbox{$D \sim 12~\mathrm{Mpc}~(E_{\mathrm{GW}}^{\mathrm{iso}}/0.01M_{\odot}c^2)^{1/2}$} 
for emission at frequencies around 150 Hz, where the LIGO-Virgo detector 
network has best sensitivity.   
We present astrophysical interpretations and implications of these results, and 
prospects for corresponding searches during future LIGO-Virgo runs. 

\end{abstract}

\keywords{gamma-ray bursts -- gravitational waves -- compact object mergers -- soft gamma-ray repeaters }

\pacs{
04.80.Nn, 
07.05.Kf, 
95.85.Sz  
97.60.Bw  
}

\section{Introduction} 
\label{sec:intro} 

Gamma-ray bursts (GRBs) are intense flashes of $\gamma$-rays which 
occur approximately once per day and are isotropically distributed 
over the sky~\cite[see, e.g.:][and references therein]{Meszaros:2006rc}.
The variability of the bursts on time scales as short as a 
millisecond indicates that the sources are very compact, while the 
identification of host galaxies and the measurement of redshifts for 
more than 100 bursts have shown that GRBs are of extra-galactic origin.  

GRBs are grouped into two broad classes by their characteristic duration 
and spectral hardness~\citep{ck93,Gehrels2006}.  The progenitors of most 
short GRBs ($\lesssim$ 2 s, with hard spectra) are widely thought to be 
mergers of neutron star binaries or neutron star-black hole binaries; 
see for example \citet{nakar-2007,PhysRevD.77.084015,PhysRevD.78.024012,PhysRevLett.100.191101,PhysRevD.79.044024}.
A small fraction (up to 
$\simeq$15\%) of short-duration GRBs are also thought to be due to giant 
flares from a local distribution of soft-gamma repeaters (SGRs) 
\citep{duncan92,2005Natur.438..991T,NaGaFo:06,2009MNRAS.395.1515C}.  Long GRBs ($\gtrsim$ 2 s, 
with soft spectra), on the other hand, are associated with core-collapse 
supernovae \citep{galama98,hjorth03,Ma_etal:04,Campana:2006qe}.  
Both the merger and supernova scenarios result in the formation of a 
stellar-mass black hole with accretion disk \citep{fryer99,Cannizzo:2009qv}, 
and the emission of gravitational radiation is expected in this process.  

To date, several searches for gravitational-wave bursts (GWBs) associated with 
gamma-ray bursts (GRBs) have been performed using data from LIGO or Virgo.  
Data from the second LIGO science run were used to search for a 
gravitational-wave signal from GRB~030329/SN~2003dh \citep{abbottgrb05}, a bright 
GRB and associated supernova located at a redshift of $z=0.1685$.
This was followed by a search for GWBs coincident with 39 GRBs which were 
detected during the second, third, and fourth LIGO science runs \citep{multigrb07}.  
Data from the Virgo detector were used to search for a GWB associated with 
GRB 050915a \citep{Ac_etal:07,Ac_etal:08}.  
Most recently, data from the fifth LIGO science run was analyzed to search 
for a GWB or binary coalescence inspiral signal from GRB~070201 \citep{grb070201_07}.  
This short-duration GRB 
had a position error box overlapping 
the Andromeda galaxy (M31), located at a distance of 770 kpc. 
No evidence for a gravitational-wave signal was found in these searches.  
In the case of GRB~070201, the non-detection of associated gravitational waves provided 
important information about its progenitor, ruling out a compact-object binary 
in M31 with high confidence.

In this paper, we present the results of a search for gravitational-wave 
bursts associated with 137 GRBs that were detected by satellite-based gamma-ray 
experiments during the fifth LIGO science run (S5) and first Virgo science run 
(VSR1), which collectively spanned the period from 2005 November 4 to 
2007 October 1.  This is the first joint search for gravitational waves by LIGO and Virgo; 
it also uses improved methods compared to previous searches, and is thus able 
to achieve better sensitivity.  

We search for GWBs from both short- and long-duration GRBs. 
Since the precise nature of the radiation depends on the somewhat-unknown progenitor model, 
and we analyse both short and long GRBs, the search methods presented in 
this paper do not require specific knowledge of the gravitational waveforms. 
Instead, we look for unmodelled burst signals with duration $\lesssim1$ s 
and frequencies in the LIGO/Virgo band, approximately 60 Hz $-$ 2000 Hz.  
The results of a template-based 
search specifically targeting binary inspiral gravitational-wave signals 
associated with short GRBs are presented separately \citep{CBCGRB}.

Although it is expected that most GRB progenitors will be at distances too 
large for the resulting gravitational-wave signals to be detectable by LIGO and Virgo  
\citep{berger05}, it is possible that a few GRBs could be located nearby.
For example, the smallest observed redshift of an optical GRB afterglow 
is $z=0.0085$ ($\simeq36$Mpc), for GRB 980425 \citep{kulkarni98,galama98,iwamoto98}; 
this would be within the LIGO-Virgo detectable range for some progenitor models.  
Recent studies \citep{2007ApJ...662.1111L,2007MNRAS.382L..21C} indicate the 
existence of a local population of under-luminous long GRBs with an observed 
rate density (number per unit volume per unit time) 
approximately $10^3$ times that of the high-luminosity population.
Also, observations seem to suggest that short-duration GRBs tend to have smaller 
redshifts than long GRBs \citep{GuPi:05,fox05}, and this has led to fairly 
optimistic estimates \citep{NaGaFo:06,GuPi:06,GuSt:09,LeFrJoMaMaSu:09} for detecting associated 
gravitational-wave emission in an extended LIGO science run.
Approximately 70\% of the GRBs in our sample do not have measured redshifts, 
so it is possible that one or more could be much closer than the typical Gpc 
distance of GRBs.

The paper is organized as follows.  Section~\ref{sec:s5run} describes 
the LIGO and Virgo detectors, and Sec.~\ref{sec:grbsample} describes 
the GRB sample during LIGO Science Run 5 / Virgo Science Run 1.  
We summarize the analysis procedure in Sec.~\ref{sec:params}. 
Two independent analysis ``pipelines'' are used to search for GWBs.  
Section~\ref{sec:results} details the results of the search.  No 
significant signal is found in association with any of the 137 GRBs studied. 
A statistical analysis of the collective GRB sample also shows no sign 
of a collective signature of weak GWBs.  
In Sec.~\ref{sec:limits} 
we place upper limits on the amplitude of gravitational waves associated with 
each GRB.  We also set lower limits on the distance to each GRB assuming 
a fixed energy emission in gravitational waves.  We conclude in 
Sec.~\ref{sec:conclusion} with some comments on the astrophysical 
significance of these results and the prospects for future GRB searches.

\section{LIGO Science Run 5 \& Virgo Science Run 1}
\label{sec:s5run}

The LIGO detectors are kilometer-scale power-recycled 
Michelson interferometers with orthogonal Fabry-Perot arms 
\citep{abbottnim04,abbott-2007}.  They are designed to 
detect gravitational waves with frequencies ranging from 
$\sim40$~Hz to several kHz.  The interferometers' maximum 
sensitivity occurs near 150~Hz.  There are two LIGO observatories: one 
located at Hanford, WA and the other at Livingston, LA.  The Hanford site 
houses two interferometers: one with 4 km arms (H1), and the other with 2 km arms 
(H2).  The Livingston observatory has one 4 km interferometer (L1).  The 
observatories are separated by a distance of 3000~km, corresponding to a
time-of-flight separation of 10~ms.

The Virgo detector (V1) is in Cascina near Pisa, Italy.  
It is a 3 km long power-recycled Michelson interferometer with orthogonal 
Fabry-Perot arms \citep{virgo08}.  During VSR1, the Virgo detector had sensitivity similar 
to the LIGO 4 km interferometers above approximately 500 Hz.  
The time-of-flight separation between the Virgo and Hanford observatories 
is 27 ms, and between Virgo and Livingston it is 25 ms.

A gravitational wave is a spacetime metric perturbation that is manifested 
as a time-varying quadrupolar strain, with two polarization components.
Data from each interferometer record the length difference of the 
arms and, when calibrated, measure the strain induced by a gravitational 
wave.  These data are in the form of a time series, digitized at a 
sample rate of 16384 s$^{-1}$ (LIGO) or 20000 s$^{-1}$ (Virgo).  
The response of an interferometer to a given strain is measured by injecting sinusoidal 
excitations with known amplitude into the test mass control systems and 
tracking the resulting signals at the measurement point throughout each run.  
The result is a measurement of the time-varying, frequency-dependent 
response function of each interferometer.  

\begin{figure}
\includegraphics[width=0.45\textwidth]{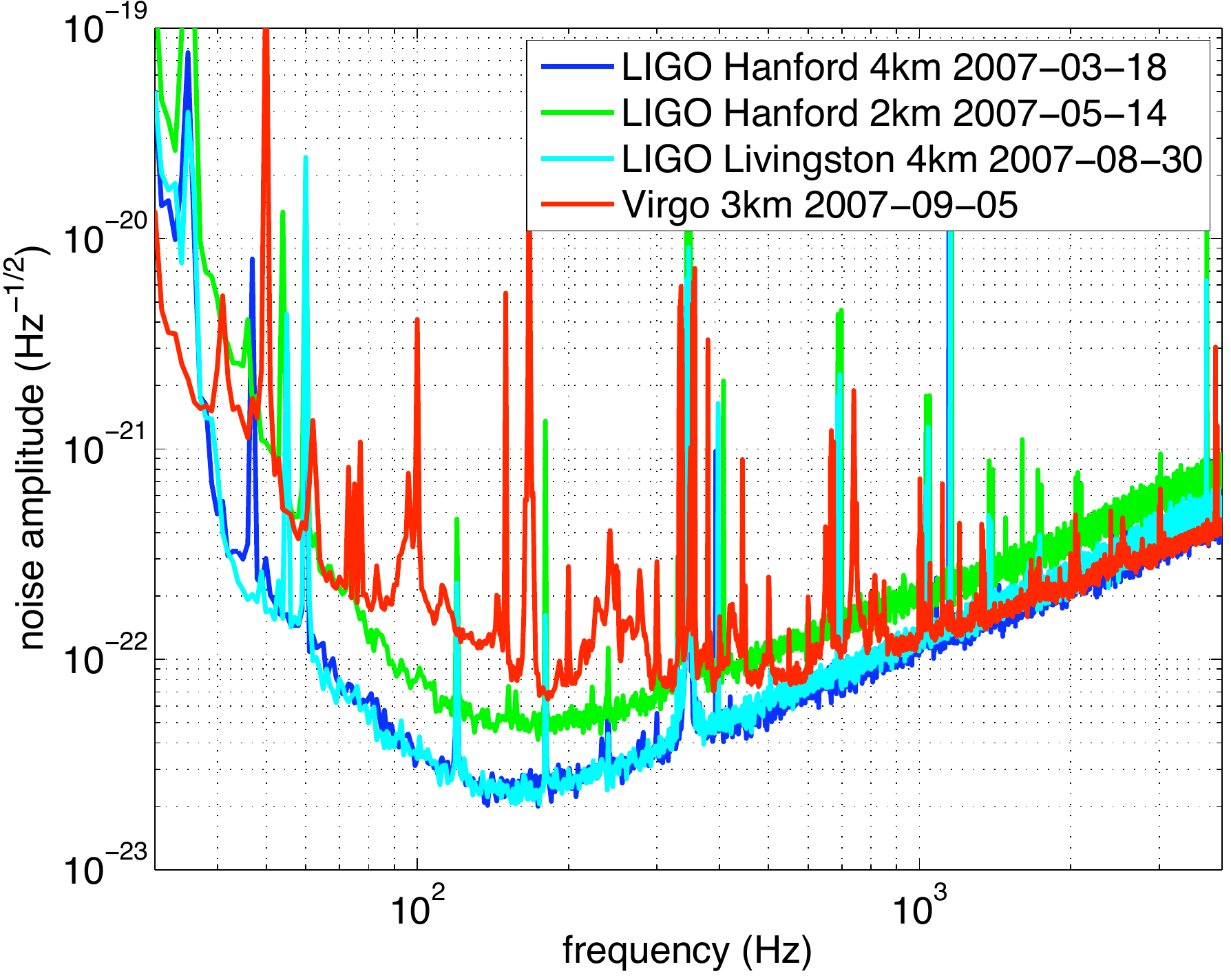}
\caption{\label{fig:spectra}Best strain noise spectra from the LIGO and Virgo 
detectors during S5-VSR1.}
\end{figure}

The fifth LIGO science run (S5) was held from 2005 November 4 to 2007 October 1.  
During this run, over one year of science-quality data was collected with 
all three LIGO interferometers in simultaneous operation.  The LIGO interferometers 
operated at their design sensitivity, with duty factors of 75\%, 76\%, and 
65\% for the H1, H2, and L1 interferometers.  The Virgo detector started 
its first science run (VSR1) on 2007 May 18.  The Virgo duty cycle over VSR1 was 78\%.
Figure~\ref{fig:spectra} shows the best sensitivities, in terms of 
noise spectral density, of the LIGO and Virgo interferometers during the run.
All of the instruments ran together continuously until 2007 October 1, 
amounting to about 4.5 months of joint data taking.  

The GEO~600 detector \citep{geo08}, located near Hannover, Germany, was also 
operational during the S5-VSR1 run, though with a lower sensitivity than LIGO 
and Virgo. We do not use the GEO data in this search as the modest gains in the 
sensitivity to gravitational wave signals would not have offset the increased 
complexity of the analysis.

\section{GRB Sample}
\label{sec:grbsample}

The GRB triggers that were contemporaneous with the S5-VSR1 run came 
mostly from the {\em Swift} satellite \citep{swift04}, but several triggers 
also came from other IPN satellites \citep{ipn2009}, including HETE-2 \citep{hete03}, and INTEGRAL \citep{integral03}.  
We obtained our GRB triggers through the Gamma-ray Burst Coordinates Network \citep{GCN}.
During the S5-VSR1 run, there were a total of 212 GRBs reported 
by these satellite-based gamma-ray experiments.  Of these, 33 were 
short-duration GRBs, and 59 had associated redshift measurements.  
All but 4 of these GRBs had well-defined positions.

Only LIGO and Virgo data which are of science-mode quality are analyzed.  
These are data collected when the interferometers are in a stable, resonant 
configuration.  Additionally, data segments which are flagged as being of 
poor quality are not included in the analysis.  A full analysis (detection 
search and upper limit calculation) is performed for all GRBs which have 
well-defined positions and for which at least two interferometers have science-mode 
data passing quality requirements.  There are 137 such GRBs, of which 21 
are short-duration bursts, and 35 have measured redshifts.  A list of the 
GRBs and relevant information are given in Table~\ref{tab:combined} in 
Appendix~\ref{sec:grbs}.

\section{Search Procedure}
\label{sec:params}

\subsection{\label{sec:overview}Overview}

The basic search procedure follows that used in recent LIGO GRB searches 
\citep{grb070201_07,multigrb07}.  All GRBs are treated identically, 
without regard to their duration, redshift (if known), or fluence.  
We use the interval from 120 s before each GRB trigger time to 60 s after 
as the window in which to search for an associated gravitational-wave burst.  
This conservative window is large enough to take into account most reasonable 
time delays between a gravitational-wave signal from a progenitor and the 
onset of the gamma-ray signal.  
For example, it is much larger than the $O(10)$ s delay of the gamma-ray 
signal resulting from the sub-luminal propagation of the jet 
to the surface of the star in the collapsar model for long GRBs 
\citep[see for example][]{AlMuIbMaMa:00,ZhWoMa:03,WaMe:07,LaMoBe:09}.  
It is also much longer than the $\lesssim 1$ s delay which may occur 
in the binary neutron star merger scenario for short GRBs 
if a hypermassive neutron star is formed \citep[see for example][]{ 
PhysRevD.78.024012,Baiotti:2008ra,PhysRevD.80.064037}.
Our window is also safely larger than any uncertainty 
in the definition of the measured GRB trigger time.  The data in this search 
window are called the {\it on-source} data.

The on-source data are scanned by an algorithm designed to detect transients 
that may have been caused by a gravitational-wave burst.  In this search, two 
algorithms are used: the cross-correlation algorithm used in previous LIGO 
searches \citep{multigrb07}, and \xp\footnote{https://geco.phys.columbia.edu/xpipeline/browser?rev=2634}, a new coherent analysis package 
\citep{Ch_etal:06,Sutton:2009gi}.  
The cross-correlation algorithm correlates the data between pairs of detectors, 
while \xp~combines data from arbitrary sets of detectors, taking into account 
the antenna response and noise level of each detector to improve the search 
sensitivity.

The data are analysed independently by \xp~and the cross-correlation algorithm 
to produce lists of transients, or {\it events}, that may be candidate 
gravitational-wave signals.  Each event is characterised by a measure of 
significance, based on energy (\xp) or correlation between detectors 
(cross-correlation algorithm).  To reduce the effect of non-stationary 
background noise, the list of candidate events is subjected to 
checks that ``veto'' events overlapping in time with known instrumental or 
environmental disturbances \citep{S5firstyear}.  \xp~also applies additional consistency tests 
based on the correlations between the detectors to further reduce the number 
of background events.  The surviving event with the largest significance is 
taken to be the best candidate for a gravitational-wave signal for that GRB; 
it is referred to as the {\it loudest event} \citep{Brady:2004gt,Biswas:2007ni}.

To estimate the expected distribution of the loudest events under 
the null hypothesis, the pipelines are also applied to all coincident data within a  3 h period surrounding the on-source data.  This data for background estimation is called 
the {\it off-source} data.  Its proximity to the on-source data makes it likely 
that the estimated background will properly reflect the noise properties in the 
on-source segment.
The off-source data are processed identically to the 
on-source data; in particular, the same data-quality cuts and consistency tests 
are applied, and the same sky position relative to the Earth is used.  
To increase the off-source distribution statistics, multiple 
time shifts are applied to the data streams from different detector sites 
(or between the H1 and H2 streams for GRBs occurring when only those two 
detectors were operating), and the off-source data are re-analysed for each time shift.
For each 180 s segment of off-source data, the loudest surviving 
event is determined.  The distribution of significances of the loudest background 
events, $C({\cal S}_\mathrm{max})$, thus gives us an empirical measure of the 
expected distribution of the significance of the loudest on-source event 
${\cal S}_\mathrm{max}^\mathrm{on}$ under the null hypothesis.

To determine if a GWB is present in the on-source data, the loudest on-source 
event is compared to the background distribution.  If the on-source significance is larger 
than that of the loudest event in 95\% of the off-source segments (i.e., 
if $C({\cal S}_\mathrm{max}^\mathrm{on})\ge0.95$), then the event is considered 
as a candidate gravitational-wave signal.  Candidate signals are subjected to 
additional ``detection checklist'' studies to try to determine the physical 
origin of the event; these studies may lead to rejecting the event as being of 
terrestrial origin, or they may increase our degree of confidence in it being due 
to a gravitational wave.

Regardless of whether a statistically 
significant signal is present, we also set a frequentist upper limit on the 
strength of gravitational waves associated with the GRB.  For a given 
gravitational-wave signal model, we define the 90\% confidence level 
upper limit on the signal amplitude as the minimum amplitude for which 
there is a 90\% or greater chance that such a signal, if present in the 
on-source region, would have produced an event with significance larger 
than the largest value ${\cal S}_\mathrm{max}^\mathrm{on}$ actually measured.
The signal models simulated are discussed in Sec.~\ref{sec:simulations}.

Since \xp~was found to be more sensitive to GWBs than the cross-correlation 
pipeline (by about a factor of 2 in amplitude), we decided in advance 
to set the upper limits using the \xp~results.  The cross-correlation 
pipeline is used as a detection-only search.  Since it was used previously 
for the analysis of a large number of GRBs during S2-S4, and for GRB070201 
during S5, including the cross-correlation pipeline provides continuity 
with past GRB searches and allows comparison of \xp~with the technique used 
for these past searches.

\subsection{\label{sec:xp}\xp}

\xp~is a \matlab-based software package for performing coherent searches for 
gravitational-wave bursts in data from arbitrary networks of detectors.  Since 
\xp~has not previously been used in a published LIGO or Virgo search, in this 
section we give a brief overview of the main steps followed in a GRB-triggered 
search.  For more details on \xp, see \citet{Sutton:2009gi}.

Coherent techniques for GWB detection 
\citep[see for example][]{GuTi:89,FlHu:98b,AnBrCrFl:01,KlMoRaMi:05,KlMoRaMi:06,MoRaKlMi:06,Ra:06,Ch_etal:06,Summerscales:2007xq} 
combine data from multiple detectors before scanning it for candidate 
events.  They naturally take into account 
differences in noise spectrum and antenna response of the detectors in 
the network.  \xp~constructs several different linear combinations of the 
data streams: those that maximize the expected signal-to-noise ratio for a 
GWB of either polarization from a given sky position (referred to as the 
$d_+$ and $d_\times$ streams), and those in which the GWB signal cancels 
(referred to as the $d_\mathrm{null}$ streams).  It then looks for transients in the 
$d_+$ and $d_\times$ streams.  Later, the energies in the $d_+$, $d_\times$, and 
$d_\mathrm{null}$ streams are compared to attempt to discriminate between true GWBs 
and background noise fluctuations.

\subsubsection{Event Generation}
\label{sec:event}

\xp~processes data in 256 s blocks.  
First, it whitens the data from each detector using linear predictor error 
filters \citep{ChBlMaKa:04}.  It then time-shifts each stream according to the 
time-of-flight for a gravitational wave incident from the sky position of the GRB, 
so that a gravitational-wave signal will be simultaneous in all the data streams 
after the shifting.  The data are divided into 50\% overlapping segments and Fourier 
transformed.  
\xp~then coherently sums and squares these Fourier series to produce time-frequency 
maps of the energy in the $d_+$, $d_\times$, and $d_\mathrm{null}$ combinations.  
Specifically, we define the noise-weighted antenna response vectors $\Fplus$ and 
$\Fcross$ for the network, with components 
\bea
	f^{+,\mathrm{DPF}}_{\alpha}(\theta,\phi,f) 
		& = & \frac{F^{+}_\alpha(\theta,\phi,\psi^{\mathrm{DPF}})}{\sqrt{S_\alpha(f)}}   \, ,
\eea
\bea
	f^{\times,\mathrm{DPF}}_{\alpha}(\theta,\phi,f) 
		& = & \frac{F^{\times}_\alpha(\theta,\phi,\psi^{\mathrm{DPF}})}{\sqrt{S_\alpha(f)}}   \, .
\eea
Here $(\theta,\phi)$ is the direction to the GRB, $\psi^{\mathrm{DPF}}$ is 
the polarization angle specifying the orientation of the plus and cross 
polarizations, $F^+_\alpha,~F^\times_\alpha \in [-1,1]$ are the antenna 
response factors to the plus and cross polarizations 
\citep[][see also Sec.~\ref{sec:simulations}]{AnBrCrFl:01},
and $S_\alpha$ is the noise power spectrum of detector $\alpha$.  
DPF stands for the dominant polarization frame; this is a 
frequency-dependent polarization basis $\psi^{\mathrm{DPF}}(f)$ such that 
$\Fplus \cdot \Fcross = 0$ and $|\Fplus| \ge |\Fcross|$ \citep{KlMoRaMi:05}.  
With this choice of basis, the $d_+$ stream is defined as the projection
\begin{equation}\label{eqn:dp}
	d_+  \equiv  \frac{\Fplus \cdot \data}{|\Fplus|}  \, , 
\end{equation}
where $\data$ is the set of whitened data streams from the individual 
detectors.  
The ``signal energy'' $\Ep\equiv|d_+|^2$ can be shown to be the sum-squared 
signal-to-noise ratio in the network corresponding to 
the least-squares estimate of the $\hp$ polarization 
of the gravitational wave in the dominant polarization frame.
The $d_\times$ stream and 
energy $\Ec$ are defined analogously.  The sum $\Ep+\Ec$ is then the maximum 
sum-squared signal-to-noise at that frequency that is consistent with a GWB 
arriving from the given sky position at that time.

The projections of the data orthogonal to $\Fplus$, $\Fcross$ yield the 
null streams, in which the contributions of a real gravitational wave incident from the given 
sky position will cancel.  The null stream energy $\En\equiv|d_\mathrm{null}|^2$ 
should therefore be consistent with background noise.  (The definition of 
the null streams is independent of the polarization basis used.)
The number of independent data combinations yielding null streams depends 
on the geometry of the network.  Networks containing both the H1 and H2 
interferometers have one null stream combination.  Networks containing 
L1, V1, and at least one of H1 or H2 have a second null stream.  For the 
H1-H2-L1-V1 network there are two independent null streams; in this case 
we sum the null energy maps from the two streams to yield a single null energy.

Events are selected by applying a threshold to the $\Ep+\Ec$ map, so that 
the pixels with the $1\%$ highest values are marked as \emph{black pixels}.  
Nearest-neighbor black pixels are grouped together into clusters \citep{Sy:02}.  
These clusters are our events. 
Each event is assigned an approximate statistical significance ${\cal S}$ 
based on a $\chi^2$ distribution; for Gaussian noise in the absence of a signal, 
$2(\Ep+\Ec)$ is $\chi^2$-distributed with $4N_\mathrm{pix}$ degrees of freedom, where 
$N_\mathrm{pix}$ is the number of pixels in the event cluster.  This significance is 
used when comparing different clusters to determine which is the ``loudest''. 
The various coherent energies ($\Ep$, $\Ec$, $\En$) are summed over 
the component pixels of the cluster, and other properties such as 
duration and bandwidth of the cluster are also recorded.

The analysis of time shifting, FFTing, and cluster identification is repeated for 
FFT lengths of $(1/8, 1/16, 1/32, 1/64, 1/128, 1/256)$ s, to cover a range of 
possible GWB durations.  Clusters produced by different FFT lengths 
that overlap in time and frequency are compared. The cluster with the largest 
significance is kept; the others are discarded.
Finally, only clusters with central time in the on-source window of 120 s before 
the GRB time to 60 s after are considered as possible candidate events.

\subsubsection{Glitch Rejection}
\label{sec:veto}

Real detector noise contains \emph{glitches}, which are short transients 
of excess strain noise that can masquerade as GWB signals.  As shown in 
\citet{Ch_etal:06}, one can construct tests that are effective at rejecting 
glitches.  Specifically, each coherent energy $\Ep$, $\Ec$, $\En$ has a 
corresponding ``incoherent'' energy $\Ip$, $\Ic$, $\In$ that is 
formed by discarding the cross-correlation terms ($d_\alpha d_\beta^*$) 
when computing $\Ep = |d_+|^2$, etc.
For large-amplitude background noise glitches the coherent and 
incoherent energies are strongly correlated, $E\sim I \pm \sqrt{I}$.
For strong gravitational-wave signals one expects either $\Ep>\Ip$ and $\Ec<\Ic$ or 
$\Ep<\Ip$ and $\Ec>\Ic$ depending on the signal polarization content, 
and $\En<\In$.

\xp~uses the incoherent energies to apply a pass/fail test to each 
event.  A nonlinear curve is fit to the measured distribution of 
background events used for tuning (discussed below); specifically, 
to the median value of $I$ as a function of $E$.  Each event is 
assigned a measure of how far it is above or below the median: 
\be
\sigma \equiv (I - I_\mathrm{med}(E)) / I^{1/2} \, .
\ee
For $(\In,\En)$, an event is passed if $\sigma_\mathrm{null} > r_\mathrm{null}$, 
where $r_\mathrm{null}$ is some threshold.  For $(\Ip,\Ep)$ and $(\Ic,\Ec)$, the 
event passes if $|\sigma_+| > r_+$ and $|\sigma_\times| > r_\times$.  
(For the H1-H2 network, which contains only aligned interferometers, the 
conditions are $\sigma_+ < r_+$ and $\sigma_\mathrm{null} > r_\mathrm{null}$.)
An event may be tested for one, two, or all three of the pairs 
$(\In,\En)$, $(\Ip,\Ep)$, and $(\Ic,\Ec)$, depending on the GRB.
The choice of which energy pairs are tested and the thresholds 
applied are determined independently for each of the 137 GRBs.  
\xp~makes the selection automatically by comparing simulated GWBs to 
background noise events, as discussed below.  In addition, the 
criterion $\In\ge1.2\En$ was imposed for all H1-H2 GRBs, as this was found to be 
effective at removing loud background glitches without affecting simulated 
gravitational waves. 

In addition to the coherent glitch vetoes, events may also be 
rejected because they overlap {\em data quality flags} or {\em vetoes}, 
as mentioned in Sec.~\ref{sec:overview}.  
The flags and vetoes used are discussed in \citet{S5firstyear}.
To avoid excessive dead time due to poor data quality, we impose 
minimum criteria for a detector to be included in the network for a given GRB.   
Specifically, at least 95\% of the 180 s of on-source data must be free of data 
quality flags and vetoes, and all of the 6 s spanning the interval from 
-5 to +1 s around the GRB trigger must be free of flags and vetoes.

\subsubsection{Pipeline Tuning}
\label{sec:tuning}

To detect a gravitational wave, \xp~compares the largest significance of all 
events in the on-source time after application of vetoes, ${\cal S}_\mathrm{max}^\mathrm{on}$, 
to the cumulative distribution $C({\cal S}_\mathrm{max})$ of loudest significances 
measured in each off-source segment.  If $C({\cal S}_\mathrm{max}^\mathrm{on})\ge0.95$, 
we consider the event for follow-up study.

To maximize the sensitivity of \xp, we tune the coherent glitch 
test thresholds $r_{+}, r_{\times}, r_\mathrm{null}$ for each GRB 
to optimize the trade-off between glitch rejection and signal 
acceptance.  We do this using the off-source data and data containing 
simulated GWB signals ({\em injections}, discussed in Sec.~\ref{sec:simulations}), but not the 
on-source data.  This {\em blind} tuning avoids the 
possibility of biasing the upper limit.

The procedure is simple.  The off-source segments and injections are 
divided randomly into two equal sets: half for tuning, and half for 
sensitivity and background estimation.  
Each of $r_{+}, r_{\times}$, and $r_\mathrm{null}$ 
are tested with trial thresholds of $[0,0.5,1,1.5,\ldots,5]$, 
where a value of $0$ is treated as not testing that energy type.
For each of the $11^3=1331$ possible combinations of trial thresholds, 
the loudest surviving event in each tuning off-source segment is found.  
The injection amplitude required for 90\% of the injections to be louder 
than the $95^\mathrm{th}$ percentile of ${\cal S}_\mathrm{max}$ is computed 
for each waveform type.  The set of thresholds giving the lowest required 
injection amplitude over all waveforms is selected as optimal (at least 
one of $r_{+}, r_{\times}$, and $r_\mathrm{null}$ is required to be non-zero).  
To get an unbiased estimate of the expected sensitivity and background, we apply 
the tuned vetoes to the second set of off-source segments and injections, 
that were not used for tuning.  For more details, see \citet{Sutton:2009gi}.

\subsection{\label{sec:cc}Cross-Correlation Pipeline}

The cross-correlation pipeline has been used in two previous 
LIGO searches \citep{multigrb07,grb070201_07} for GWBs associated 
with GRBs, and is described in detail in these references.  We 
therefore give only a brief summary of the pipeline here.  

In the present search, the cross-correlation pipeline is applied 
to the LIGO detectors only.  (The different orientation and noise 
spectrum shape of Virgo relative to the LIGO detectors is more easily 
accounted for in a coherent analysis.)
The 180 s on-source time series for each interferometer is whitened as described in 
\citet{multigrb07} and divided into time bins, then the 
cross-correlation for each interferometer pair and time 
bin is calculated.  The cross-correlation $cc$ of two 
timeseries $s_1$ and $s_2$ is defined as 
\begin{equation}
cc = \frac{\sum_{i=1}^m [s_1(i)-\mu_1][s_2(i)-\mu_2]}{
         \sqrt{\sum_{j=1}^m [s_1(j)-\mu_1]^2}
         \sqrt{\sum_{k=1}^m [s_2(k)-\mu_2]^2 \vphantom{\sum_{j=1}^m}}
     } \, ,
\end{equation}
where $\mu_1$ and $\mu_2$ are the corresponding means, 
and $m$ is the number of samples in the bin. 
Cross-correlation bins of lengths 25 ms and 100 ms are 
used to target short-duration GW signals with durations of 
$\sim1$ ms to $\sim100$ ms.
The bins are overlapped by half a bin width to avoid 
loss of signals occurring near a bin boundary. 
Each LHO-LLO pair of 180 s on-source segments is shifted 
in time relative to each other to account for the time-of-flight 
between the detector sites for the known sky position of the GRB 
before the cross-correlations are calculated. 

The cross-correlation is calculated for each interferometer pair and time bin 
for each bin length used. For an H1-H2 search the largest 
cross-correlation value obtained within the 180 s  
search window is considered the most significant 
measurement. For an H1-L1 or 
H2-L1 search, the largest {\it absolute} value of the 
cross-correlation is taken as the most significant 
measurement. This was done to take into account the 
possibility that signals at LHO and LLO could be 
anticorrelated depending on the (unknown) polarization 
of the gravitational wave.

\section{Search Results}
\label{sec:results}

\subsection{Per-GRB Results}

The results of the search for each of the 137 GRBs analysed by 
\xp~are shown in Table~\ref{tab:combined}, Appendix~\ref{sec:grbs}.  
The seventh column in this table lists the {\em local probability} 
$p \equiv 1-C({\cal S}_\mathrm{max}^\mathrm{on})$ for the loudest on-source 
event, defined as the fraction of background trials that produced 
a more significant event (a ``$-$'' indicates no event survived all cuts).
Five GRBs had events passing the threshold of $p=0.05$ to become 
candidate signals. 

Since the local probability is typically estimated using approximately 
150 off-source segments, small $p$ values are subject to relatively large 
uncertainty from Poisson statistics.  We therefore applied additional 
time-shifts to the 
off-source data to obtain a total of 18000 off-source segments for each 
candidate which were processed to improve the 
estimate of $p$.  The 5 GRBs and their refined local probabilities are
060116 ($p=0.0402$), 060510B (0.0124), 060807 (0.00967), 061201 (0.0222), 
and 070529 (0.0776).  (Note that for GRB 070529, the refined local 
probability from the extra off-source segments was {\em larger} than the 
threshold of 0.05 for candidate signals.)  

Considering that we analysed 137 GRBs, these numbers 
are consistent with the null hypothesis that no gravitational-wave burst 
signal is associated with any of the GRBs tested.
In addition, three of these GRBs have large measured redshift: 
GRB 060116 ($z=6.6$), 060510B ($z=4.9$), and 070529 ($z=2.5$), making 
it highly unlikely {\em a priori} that we would expect to see 
a GWB in these cases.  
Nevertheless, each event has been subjected to follow-up examinations.  
These include checks of the consistency of the candidate with background 
events (such as in coherent energies, and frequency), checks of detector 
performance at the time as indicated by monitoring programs and operator 
logs, and scans of data from detector and environmental monitoring equipment 
to look for anomalous behavior.
In each case, the candidate event appears consistent with the 
coherent energy distributions of background events, lying just 
outside the coherent glitch rejection thresholds.  The frequency 
of each event is also typical of background events for their 
respective GRBs.  
Some of these GRBs occurred during periods of elevated 
background noise, and one occurred during a period of glitchy data in H1.
In two cases scans of data from monitoring equipment indicate a possible 
physical cause for the candidate event: one from non-stationarity in a 
calibration line, and another due to upconversion of low-frequency noise 
in H1.

All but two of the GRBs processed by \xp~are also analysed by 
the cross-correlation pipeline.  The cross-correlation pipeline 
produces a local probability for each detector pair and for each 
bin length (25 ms and 100 ms), for a total of 646 measurements from 135 GRBs.
The threshold on the cross-correlation local probability corresponding 
to the 5\% threshold for \xp~is therefore 5\%$\times135/646\simeq1\%$.
A total of 7 GRBs have $p<1\%$ from cross-correlation:
060306 (0.00833),
060719 (0.00669),
060919 (0.00303),
061110 (0.00357),
070704 (0.00123), and 
070810 (0.00119).
These results are also consistent with the null hypothesis.  
Furthermore, none of these GRBs are among those that had 
a low $p$ value from \xp.  
This is further indication that the candidate events detected by each 
pipeline are due to background noise rather than GWBs.  Specifically, 
\xp~and the cross-correlation pipeline use different measures of 
significance of candidate events.  Whereas a strong GWB should be 
detected by both, any given background noise fluctuation may have 
very different significance in the two pipelines.

We conclude that we have not identified a plausible gravitational-wave 
burst signal associated with any of the 137 GRBs tested.

\subsection{Binomial Test}

Gravitational-wave signals from individual GRBs are likely to be 
very weak in most cases due to the cosmological distances involved. 
Therefore, besides searching for GWB signals from each GRB, 
we also test for a cumulative signature associated with a sample 
of several GRBs \citep{FMR}.  This approach has been used in 
\citet{astone1,astone2} to analyze resonant 
mass detector data using triggers from the BATSE and 
BeppoSAX missions, and more recently in the LIGO search 
for GRBs during the S2, S3, and S4 science runs \citep{multigrb07}.

Under the null hypothesis (no GWB signal), the local probability 
for each GRB is expected to be uniformly distributed on $[0,1]$.  
Moderately strong GWBs associated with one or more of the GRBs 
will cause the low-$p$ tail of the distribution to deviate from 
that expected under the null hypothesis.  We apply the binomial test 
used in \citet{multigrb07} to search for a statistically significant 
deviation, applying it to the $5\% \times 137 \simeq 7$ least 
probable (lowest $p$) on-source results in the GRB distribution. 
Briefly, we sort the 7 smallest local probabilities 
in increasing order, $p_1, p_2, \ldots, p_7$.  For each $p_i$ we 
compute the binomial probability $P_{\ge i}(p_i)$ of getting $i$ 
or more events out of 137 at least as significant as $p_i$.  The 
smallest $P_{\ge i}(p_i)$ is selected as the most significant 
deviation from the null hypothesis.  To account for the trials 
factor from testing 7 values of $i$, we repeat the test many 
times using 137 fake local probabilities drawn from uniform 
discrete distributions corresponding to the number of off-source  
segments for each GRB (18000 for our refined $p$ estimates).
The probability 
associated with the actual smallest $P_{\ge i}(p_i)$ is defined as 
the fraction of Monte Carlo trials that gave binomial probabilities 
as small or smaller.  Note that this procedure also automatically 
handles the case of a single loud GWB.

\begin{figure}
\includegraphics[width=0.45\textwidth]{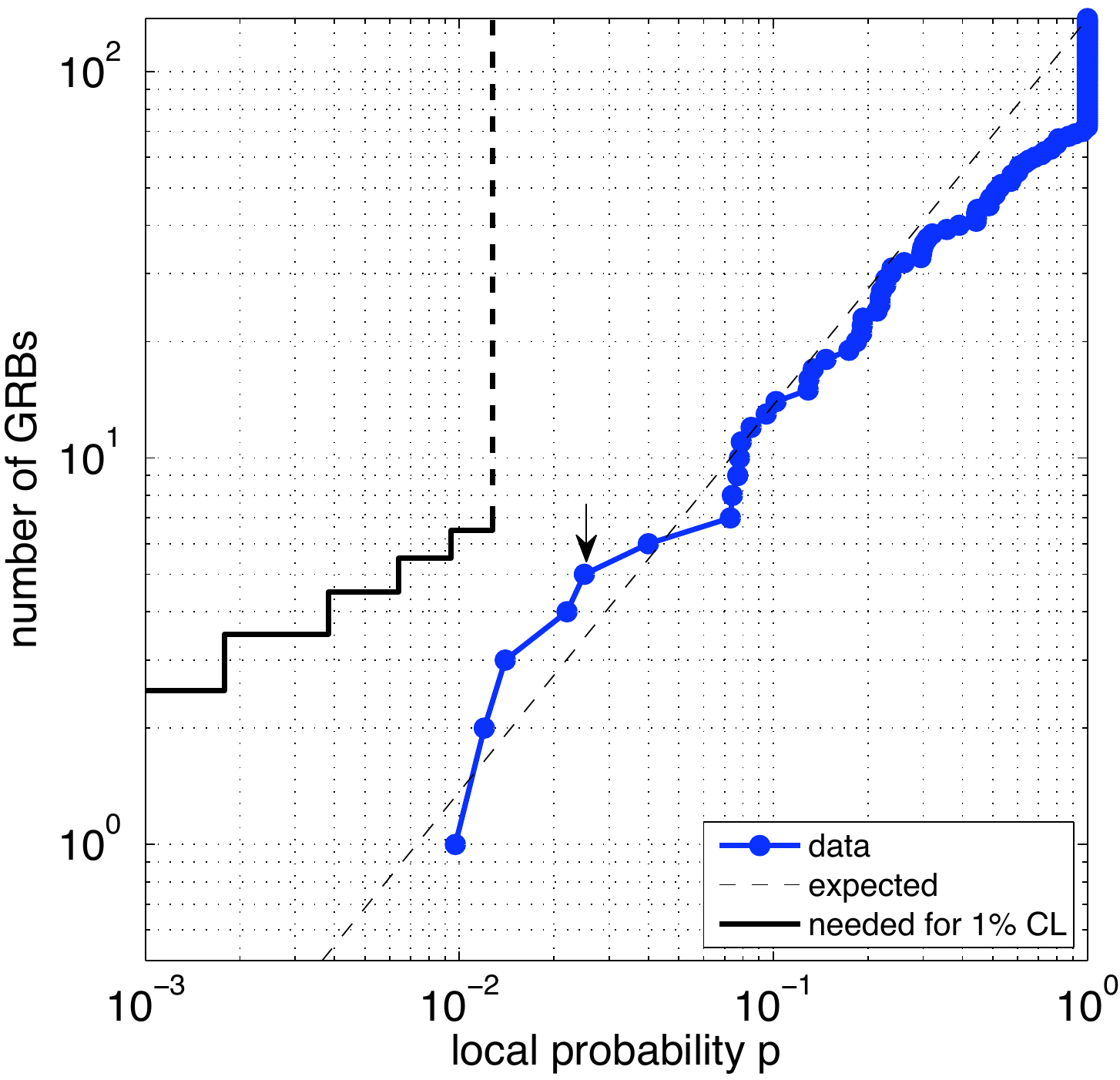}
\caption{\label{fig:binomial}
Cumulative local probability distribution resulting from the search 
of 137 GRBs with \xp.  The most significant excess is indicated 
by the arrow. The expected distribution under the null hypothesis 
is indicated by the diagonal dashed line. The excess needed for a 
1\% confidence in the null hypothesis is indicated by the solid 
line. The maximum excess indicated by this line is 7 events 
because only the 7 most significant events in the actual 
distribution are tested. The buildup of GRBs at $p=1$ occurs 
because approximately half of the GRBs do not have any event 
surviving all the analysis cuts.}
\end{figure}

In addition to the 5 ``candidate'' GRBs, extra time-shifted off-source 
segments were analysed for the 2 GRBs with the next smallest local 
probabilities, GRB 060428B ($0.0139$) and 060930 ($p=0.0248$).  
(By chance, for both of these GRBs the refined local probabilities 
from the extra off-source segments are {\em smaller} than the 
threshold of 0.05 for candidate signals, though the original estimates 
were larger.) 
Together with the 5 candidates, this gives the 7 refined local 
probabilities 0.00967, 0.0124, 0.0139, 0.0222, 0.0248, 0.0402, 0.0776.
The associated smallest binomial probability is $P_{\ge 5}(0.0248)=0.259$.  
Approximately 56\% of Monte Carlo trials give binomial probabilities 
this small or smaller, hence we conclude that there is no significant 
deviation of the measured local probabilities from the null hypothesis.  
For comparison, Figure~\ref{fig:binomial} shows the distribution of 
local probabilities for all GRBs, as well as the values that would 
need to be observed to give only 1\% consistency with the null hypothesis.

Similar results are found when restricting the test to GRBs without 
measured redshift.  In this case the smallest binomial probability is 
$P_{\ge 4}(0.0248)=0.252$ with 48\% of Monte Carlo trials yielding 
binomial probabilities this small or smaller.  Analysis of the 
cross-correlation local probabilities also shows no significant deviation.
Combining the local probabilities from the 25 ms and 100 ms analyses, 
we find the smallest binomial probability to be 
$P_{\ge 2}(0.00123)=0.190$ with 52\% of Monte Carlo trials yielding 
binomial probabilities this small or smaller. 
 
\section{Upper Limits}
\label{sec:limits}

The sensitivity of the search to gravitational waves is determined by a 
Monte Carlo analysis.  For each GRB, we add (or ``inject'') simulated GWB 
signals into the detector data and repeat the analysis.  We count an 
injected signal as ``detected'' if it produces an event that is louder 
than the loudest on-source event within 100 ms of the injection time.  
(When tuning, we do not know the significance of the loudest on-source event. 
We therefore count an injection as 
detected if it is louder than the median background loudest event from the 
off-source tuning segments; {\em i.e.}, louder than the $50^\mathrm{th}$ 
percentile of the sample of ${\cal S}_\mathrm{max}$ values.)
For a given waveform 
morphology, we define the 90\% confidence level upper limit on the signal 
amplitude as the minimum amplitude for which the detection probability 
is 0.9 or greater.

We discuss the signal models in Sec.~\ref{sec:simulations}, 
their systematic uncertainties in Sec. 6.2, and the 
upper limit results in Sec.~\ref{subsec:hrssul}.

\subsection{Simulations}
\label{sec:simulations}

The antenna response of an interferometer to a gravitational wave 
with polarization strains $h_+(t)$ and $h_\times(t)$ depends on the  
polarization basis angle $\psi$ and the direction $(\theta,\phi)$ to 
the source as
\begin{equation}
h(t) = F_+(\theta,\phi,\psi)h_+(t) + F_{\times}(\theta,\phi,\psi)h_\times(t) \, .
\label{eq:hresponse}
\end{equation}
Here $F_+(\theta,\phi,\psi)$, $F_\times(\theta,\phi,\psi)$ are the 
plus and cross antenna factors introduced in Sec.~\ref{sec:xp};
see \citet{AnBrCrFl:01} for explicit definitions.  

A convenient measure of the gravitational-wave amplitude is the root-sum-square 
amplitude,
\begin{equation}
h_{\rm rss} = \sqrt{\int (|h_+(t)|^2 + |h_\times(t)|^2) ~ dt} ~ .
\label{eq:hrss}
\end{equation}
The energy flux (power per unit area) of the wave is \citep{isaacson}
\begin{equation}
F_{\rm GW} = \frac{c^3}{16\pi G} \langle (\dot{h}_{+})^2 + (\dot{h}_{\times})^2 \rangle \, ,
\end{equation}
where the angle brackets denote an average over several periods. 
The total energy emitted assuming isotropic emission is then
\begin{equation}
E_{\mathrm{GW}}^{\mathrm{iso}} = 4\pi D^2 \int dt \, F_{\rm GW} \, ,
\end{equation}
where $D$ is the distance to the source.\

The forms of $h_+(t)$ and $h_\times(t)$ depend on the type of 
simulated waveform.  It is likely that many short GRBs are produced 
by the merger of neutron-star--neutron-star or black-hole--neutron-star 
binaries.  The gravitational-wave signal from inspiralling binaries 
is fairly well understood \citep{lrr-2006-4,Aylott:2009ya}.  Progress 
is being made on modelling the merger phase; recent numerical studies 
of the merger of binary neutron star systems and gravitational-wave 
emission have been performed by 
\citet{PhysRevD.71.084021,PhysRevD.73.064027,Baiotti:2008ra,BaGiRe:09,PhysRevD.78.064054,PhysRevD.79.124033,PhysRevD.80.064037,Rezzolla:2010fd}. 
Preliminary explorations of the impact of magnetic fields have also been made by 
\citet{PhysRevLett.100.191101,PhysRevD.78.024012,Giacomazzo:2009mp}.
The merger of black-hole--neutron-star binaries have been studied by 
\citet{PhysRevD.74.121503,ShUr:07,PhysRevD.77.084015,PhysRevD.77.084002,PhysRevD.78.104015,PhysRevD.78.064054,PhysRevD.79.044030,PhysRevD.79.044024,Duez:2009yy}. 
For other progenitor types, particularly for long GRBs, there are no robust 
models for the gravitational-wave emission 
\citep[see for example][for possible scenarios]{Fryer,kobayashi-2003-585,vanPutten:grb,Ott:2008wt}.  
Since our detection algorithm is designed 
to be sensitive to generic gravitational-wave bursts, we 
choose simple {\it ad hoc} waveforms for tuning and testing. 
Specifically, we use sine-Gaussian and cosine-Gaussian waveforms:
\begin{eqnarray}
h_+(t+t_0) &=\, h_{+,0} \sin(2 \pi f_0 t) \exp\left(\dfrac{-(2\pi f_0 t)^2}{2Q^2}\right) \,, \label{eqn:sg} \\
h_{\times}(t+t_0) &=\, h_{\times,0} \cos(2 \pi f_0 t) \exp\biggl(\dfrac{-(2\pi f_0 t)^2}{2Q^2}\biggr) \,,
\label{eqn:cg}
\end{eqnarray}
where $t_0$ is the central time, $f_0$ is the central frequency, 
$h_{+,0}$ and $h_{\times,0}$ are the amplitude parameters of the 
two polarizations, and $Q$ is a dimensionless constant which
represents roughly the number of cycles with which the waveform oscillates with
more than half of the peak amplitude.  For $Q \gtrsim 3$, the 
root-sum-squared amplitude of this waveform is
\begin{equation}
h_{\rm rss} 
  \approx  \sqrt{\dfrac{Q(h_{+,0}^2 + h_{\times,0}^2)}{4 \pi^{1/2} f_0}} 
\label{eq:rssdef1}
\end{equation}
and the energy in gravitational waves is 
\begin{equation}\label{eqn:EGW}
E_{\mathrm{GW}}^{\mathrm{iso}} 
  \approx  \frac{\pi^2c^3}{G} D^2 f_0^2  h_{\rm rss}^2 \,.
\end{equation}
Using these waveforms for $h_+(t)$ and $h_\times(t)$, we simulate  
circularly polarized GW waves by setting the sine-Gaussian and 
cosine-Gaussian amplitudes equal to each other, $h_{+,0} = h_{\times,0}$.  
To simulate linearly polarized waves, we set $h_{\times,0} = 0$.

The peak time of the simulated signals is distributed uniformly 
through the on-source interval.  
We use $Q=2^{3/2}\pi=8.9$, a standard choice in LIGO burst searches.  
The polarization angle $\psi$ for which $\hp$, $\hc$ take the 
forms in equations (\ref{eqn:sg}) and (\ref{eqn:cg}) is uniform on $[0,\pi)$, and the 
sky position used is that of the GRB (fixed in right ascension and 
declination).
We simulate signals at discrete log-spaced amplitudes, 
with 500 injections of each waveform for each amplitude.

Early tests of the search algorithms used the central frequencies 
$f_0 = (100, 150, 250, 554, 1000, 1850)$ Hz, and both linearly and 
circularly polarized injections.  The final \xp~tuning (performed after 
implementation of an improved data-whitening procedure) uses 150 Hz 
and 1000 Hz injections of both polarizations.

\subsection{Statistical and Systematic Errors}
\label{sec:errors}

Our upper limit on gravitational-wave emission by a GRB is 
$h^{90\%}_\mathrm{rss}$, the amplitude at which         
there is a 90\% or greater chance that such a signal, if present in 
the on-source region, would have produced an event with 
signiÞcance larger than the largest actually measured.
There are several sources of error, both statistical and systematic, 
that can affect our limits.  These 
are calibration uncertainties (amplitude and phase response of the 
detectors, and relative timing errors), uncertainty in the sky 
position of the GRB, and uncertainty in the measurement of 
$h^{90\%}_\mathrm{rss}$ due to the finite number of injections and 
the use of a discrete set of amplitudes. 

To estimate the effect of these errors on our upper limits, we repeat  
the Monte Carlo runs for a subset of the GRBs, simulating all three of 
these types of errors.  Specifically, the amplitude, phase, and time 
delays for each injection in each detector are perturbed by 
Gaussian-distributed corrections matching the calibration uncertainties 
for each detector.  The sky position is perturbed in a random direction 
by a Gaussian-distributed angle with standard deviation of 3 arcmin.  
Finally, the discrete amplitudes used are offset by those in the standard analysis 
by a half-step in amplitude.  The perturbed injections are then 
processed, and the open-box upper limit produced using the same coherent 
consistency test tuning as in the actual open-box search.
The typical difference between the upper limits for perturbed injections 
and unperturbed injections then gives an estimate of the impact of the 
errors on our upper limits.

For low-frequency injections (at 150 Hz) we find that the ratio of 
the upper limit for perturbed injections to unperturbed injections is 
1.03 with a standard deviation of 0.06.  We therefore increase the 
estimated upper limits at 100 Hz by a 
factor of $1.03 + 1.28\times0.06 = 1.10$ as a 
conservative allowance for statistical and systematic errors (the 
factor 1.28 comes from the 90\% upper limit for a Gaussian distribution). 
The dominant contribution is due to the finite number of injections.
For the high-frequency (1000 Hz) injections 
the factor is $1.10 + 1.28\times0.12 = 1.25$.  In addition to 
finite-number statistics, the calibration uncertainties are more 
important at high frequencies and make a significant contribution 
to this factor.  All limits reported in this paper include these 
allowance factors.

\subsection{Limits on Strain and Distance}
\label{subsec:hrssul}

The upper limits on GWB amplitude and lower limits on the distance for 
each of the GRBs analysed are given in Table~\ref{tab:combined} in 
Appendix~\ref{sec:grbs}.  These limits are computed for circularly 
polarized 150 Hz and 1000 Hz sine-Gaussian waveforms.  We compute the 
distance limits by assuming the source emitted 
$E_{\mathrm{GW}}^{\mathrm{iso}} = 0.01 M_\odot c^2 = 1.8\times10^{52}$erg 
of energy isotropically in gravitational waves and use equation (\ref{eqn:EGW}) to 
infer a lower limit on $D$. 
We choose $E_{\mathrm{GW}}^{\mathrm{iso}} = 0.01 M_\odot c^2$ because this 
is a reasonable value one might expect to be emitted in the LIGO-Virgo band 
by various progenitor models.  For example, mergers of neutron-star binaries 
or neutron-star--black-hole binaries (the likely progenitors of most short 
GRBs) will have isotropic-equivalent emission of order (0.01 $-$ 0.1) 
$M_{\odot}c^2$ in the 100-200 Hz band.  For long GRBs, 
fragmentation of the accretion disk 
\citep{2002ApJ...579L..63D,2005ApJ...630L.113K,Piro:2006ja} could produce 
inspiral-like chirps with (0.001 $-$ 0.01) $M_{\odot}c^2$ emission.  The 
suspended accretion model \citep{vanPutten:grb} also predicts an energy 
emission of up to (0.01 $-$ 0.1$) M_{\odot}c^2$ in this band.  For other 
values of $E_{\mathrm{GW}}^{\mathrm{iso}}$ the distance limit scales as 
$D \propto (E_{\mathrm{GW}}^{\mathrm{iso}})^{1/2}$.

\begin{figure}
\includegraphics[width=0.45\textwidth]{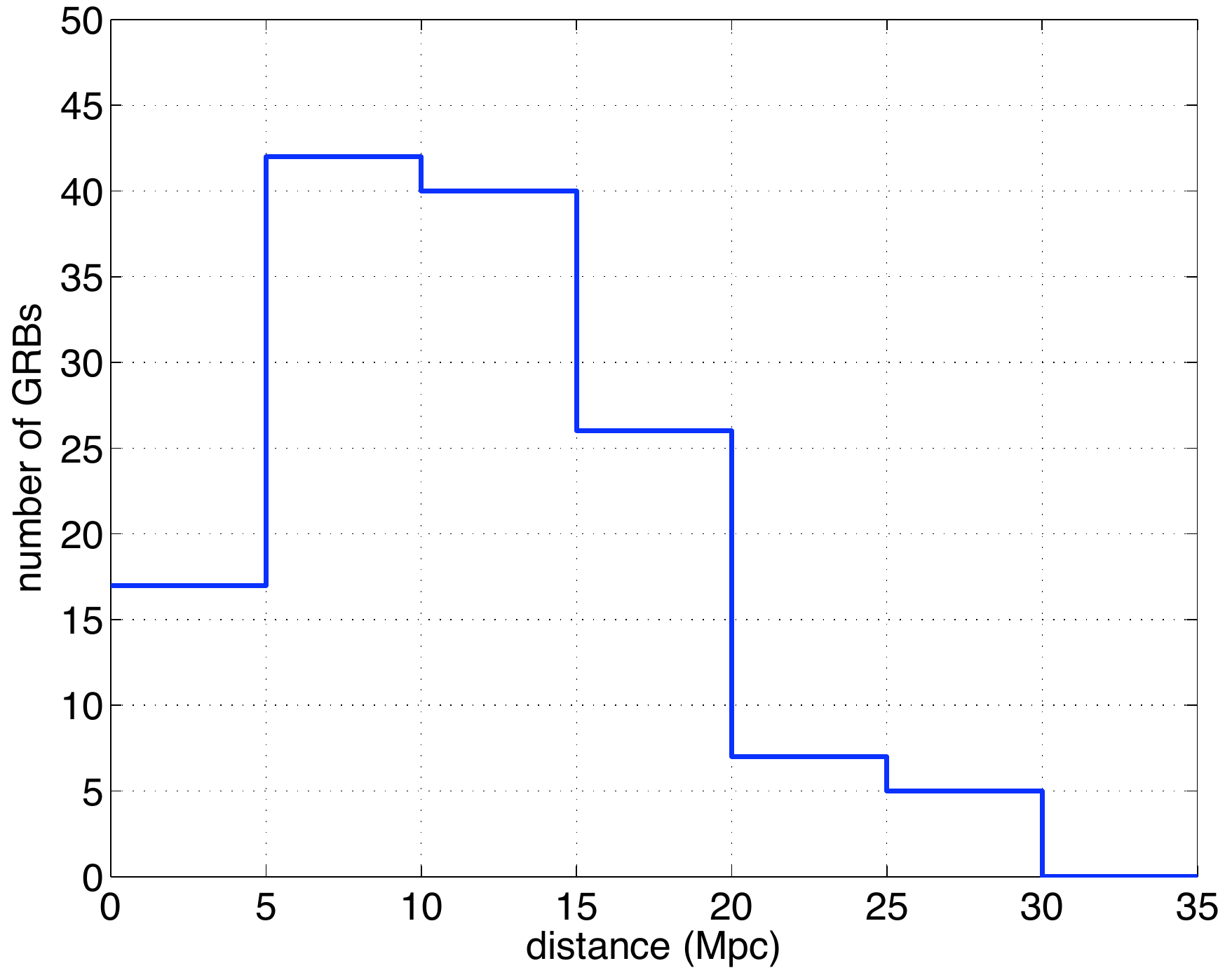}
\caption{\label{fig:Egw}Histogram of lower limits on the distance 
to each of the 137 GRBs studied, assuming that the 
GRB progenitors emit $0.01 M_{\odot}c^2 = 1.8\times10^{52}\mathrm{erg}$ 
of energy in circularly polarized gravitational waves at 150 Hz.}
\end{figure}

As can be seen from Table~\ref{tab:combined}, the strongest limits are 
on gravitational-wave emission at 150 Hz, where the sensitivity of the detectors is best 
(see Figure~\ref{fig:spectra}).  Figure~\ref{fig:Egw} shows a histogram 
of the distance limits for the 137 GRBs tested.  
The typical limits at 150 Hz from the \xp~analysis are (5$-$20) Mpc.
The best upper limits are for GRBs later in S5-VSR1, when the 
detector noise levels tended to be lowest (and when the most detectors 
were operating), and for GRBs that occurred at sky positions for which 
the detector antenna responses $F_+$, $F_\times$ were best.  
The strongest limits obtained were for GRB 070429B: 
$h^{90\%}_\mathrm{rss} = 1.75\times10^{-22}\mathrm{Hz}^{-1/2}$, 
$D^{90\%} = 26.2$ Mpc at 150 Hz.
For comparison, the smallest measured redshift in our GRB sample is 
for 060614, which had $z=0.125$ \citep{2006GCN..5275....1P} or $D\simeq578$ Mpc 
\citep{wright}. 
(Though GRB 060218 at $z=0.0331$ \citep{1538-4357-643-2-L99} occurred 
during S5, unfortunately, the LIGO-Hanford and Virgo detectors 
were not operating at the time.)

A GRB of particular interest is 070201.  This short-duration GRB had a 
position error box overlapping M31 \citep[see][and references therein]{mazets07}, 
which is at a distance of only 770 kpc.
An analysis of LIGO data from this time was presented in \citet{grb070201_07}.  
GRB 070201 was included in the present search using the new \xp~search package.  
Our new upper limits on the amplitude of a GWB associated with GRB 070201 
are $h^{90\%}_\mathrm{rss} = 6.38\times10^{-22}\mathrm{Hz}^{-1/2}$ at 150 Hz, 
and $h^{90\%}_\mathrm{rss}=27.8\times10^{-22}\mathrm{Hz}^{-1/2}$ at 1000 Hz.  
These are approximately a factor of 2 lower than those placed by the 
cross-correlation algorithm.  For a source at 770 kpc, the energy 
limit from equation (\ref{eqn:EGW}) is 
$E_{\mathrm{GW}}^{\mathrm{iso}} = 1.15\times10^{-4}M_\odot c^2$ at 150 Hz.
While about a factor of 4 lower than the GWB limit presented in 
\citet{grb070201_07}, this is still several orders of magnitude away 
from being able to test the hypothesis that this GRB's progenitor was 
a soft-gamma repeater in M31 \citep{mazets07}.

\section{Summary and Conclusion}
\label{sec:conclusion}

We have presented the results of a search for gravitational-wave bursts 
associated with 137 GRBs that occurred during the LIGO Science 
Run 5 -- Virgo Science Run 1, from 2005 November 4 to 2007 October 1.  The search 
used two independent data-analysis pipelines to scan for unmodelled 
transient signals consistent with the known time and sky position of 
each GRB.  No plausible gravitational-wave signals were identified. 
Assuming isotropic gravitational-wave emission by the progenitor, 
we place lower limits on the distance to each GRB.  The median limit is 
\mbox{$D \sim 12~\mathrm{Mpc}~(E_{\mathrm{GW}}^{\mathrm{iso}}/0.01M_{\odot}c^2)^{1/2}$} 
for emission at frequencies around 150 Hz, where the LIGO-Virgo detector 
network has best sensitivity.

It is informative to compare this result to the rate density of GRBs \citep[see, for example,][]{LeFrJoMaMaSu:09}.  
For long GRBs, a commonly used estimate of the local rate density 
(the rate of observable GRBs per unit volume) is 
$R_{\mathrm{long}}^{\mathrm{obs}}\sim 0.5~\mathrm{Gpc}^{-3}\mathrm{yr}^{-1}$ 
\citep{Sokolov01,Schmidt01,Le:2006pt}. 
We therefore estimate the {\em a priori} expected number of long GRBs 
being observed within a distance $D$ during a two-year science run as 
\begin{eqnarray}
\langle N_\mathrm{long} \rangle 
  & \simeq &  R^{\mathrm{obs}}_{\mathrm{long}} \left(\frac{4}{3}\pi D^3\right) T \frac{\Omega}{4\pi} \, , 
\end{eqnarray}
where $T$ is the total observation time with two or more gravitational-wave detectors operating, and $\Omega$ is the field of view of the satellite's GRB detector.  Most of  
the S5-VSR1 GRBs were detected by {\em Swift}, with $\Omega=1.4$ sr.  The coincident observation time was approximately 1.3 yr.  These give
\begin{eqnarray}\label{eqn:NLS5}
\langle N_\mathrm{long} \rangle 
  & \simeq &  1\times10^{-6} \frac{R_{\mathrm{long}}^{\mathrm{obs}}}{0.5~\mathrm{Gpc}^{-3}\mathrm{yr}^{-1}} \left(\frac{E_{\mathrm{GW}}^{\mathrm{iso}}}{0.01M_{\odot}c^2}\right)^{3/2} \!\!\!\!\! .  \qquad 
\end{eqnarray}

Recent studies \citep{2007ApJ...662.1111L,2007MNRAS.382L..21C} have indicated that there exists a local population of under-luminous long GRBs with an observed rate density approximately $10^3$ times that of the high-luminosity population.  For this population 
we have
\begin{eqnarray}\label{eqn:NlocalS5}
\langle N_\mathrm{local} \rangle 
  & \simeq &  1\times10^{-3} \frac{R_{\mathrm{local}}^{\mathrm{obs}}}{500~\mathrm{Gpc}^{-3}\mathrm{yr}^{-1}} \left(\frac{E_{\mathrm{GW}}^{\mathrm{iso}}}{0.01M_{\odot}c^2}\right)^{3/2} \!\!\!\!\! .  \qquad 
\end{eqnarray}

For short GRBs the estimated local rate density is of order 
$R_{\mathrm{short}}^{\mathrm{obs}}\sim 10~$Gpc$^{-3}$yr$^{-1}$ \citep{GuPi:06,NaGaFo:06}. 
We therefore estimate the {\em a priori} expected number of short GRBs being 
observed during S5-VSR1 as 
\begin{eqnarray}\label{eqn:NSS5}
\langle N_\mathrm{short} \rangle 
  & \simeq &  2\times10^{-5} \frac{R_{\mathrm{short}}^{\mathrm{obs}}}{10~\mathrm{Gpc}^{-3}\mathrm{yr}^{-1}} \left(\frac{E_{\mathrm{GW}}^{\mathrm{iso}}}{0.01M_{\odot}c^2}\right)^{3/2} \!\!\!\!\! .  \qquad 
\end{eqnarray}
There is also evidence of a high-density local population of short GRBs 
\citep{2005Natur.438..991T,NaGaFo:06,2009MNRAS.395.1515C}, but these are thought to be due to 
extra-galactic SGRs, which are not so promising as GW sources.

It is clear that the detection of gravitational-wave emission 
associated with either a short or long GRB with the current 
LIGO-Virgo network is unlikely, though not impossible.  
Looking ahead, the enhanced LIGO and Virgo detectors have recently  
begun their next data-taking run, S6-VSR2.
Furthermore, the Fermi satellite is now operating, with a 
field of view of approximately $\Omega=9.5$ sr. 
Assuming a similar observation time and sensitivity for S6-VSR2, 
the expected number of detections scales to 
\begin{eqnarray}
\langle N_\mathrm{long} \rangle 
  & \simeq &  7\times10^{-6}  \\  
\langle N_\mathrm{local} \rangle 
  & \simeq &  7\times10^{-3}  \\ 
\langle N_\mathrm{short} \rangle 
  & \simeq &  1\times10^{-4} \, , 
\end{eqnarray}
where we use the nominal values for $R^{\mathrm{obs}}$, 
$E_{\mathrm{GW}}^{\mathrm{iso}}$ as in equations (\ref{eqn:NLS5})--(\ref{eqn:NSS5}).
Further in the future (c.2015), the planned advanced LIGO \citep{aligo} and 
advanced Virgo \citep{avirgo} detectors 
will have amplitude sensitivities about an order 
of magnitude greater than the current detectors.  Since the 
search volume scales as $D^3$, there 
is a very good chance that we will be able to detect gravitational 
waves associated with one or more GRBs during an extended science 
run of the advanced detectors.

\acknowledgments
We are indebted to the observers of the electromagnetic events 
and the GCN for providing us with valuable data.
The authors gratefully acknowledge the support of the United States 
National Science Foundation for the construction and operation of the 
LIGO Laboratory, the Science and Technology Facilities Council of the 
United Kingdom, the Max-Planck-Society, and the State of 
Niedersachsen/Germany for support of the construction and operation of 
the GEO600 detector, and the Italian Istituto Nazionale di Fisica 
Nucleare and the French Centre National de la Recherche Scientifique 
for the construction and operation of the Virgo detector. The authors 
also gratefully acknowledge the support of the research by these 
agencies and by the Australian Research Council, the Council of 
Scientific and Industrial Research of India, the Istituto Nazionale di 
Fisica Nucleare of Italy, the Spanish Ministerio de Educaci\'on y 
Ciencia, the Conselleria d'Economia Hisenda i Innovaci\'o of the 
Govern de les Illes Balears, the Foundation for Fundamental Research 
on Matter supported by the Netherlands Organisation for Scientific 
Research, the Royal Society, the Scottish Funding Council, the 
Scottish Universities Physics Alliance, The National Aeronautics and 
Space Administration, the Carnegie Trust, the Leverhulme Trust, the 
David and Lucile Packard Foundation, the Research Corporation, and 
the Alfred P. Sloan Foundation.
This document has been assigned LIGO Laboratory document
number {LIGO-P09}{00023}-v16.

\appendix
\section{GRB Sample and Search Results}
\label{sec:grbs}

Table~\ref{tab:combined} lists the 137 GRBs analyzed in this analysis, 
including the GRB name, time, sky position, and redshift (when known).   
In addition, for each GRB we display the results of the \xp~search for 
an associated GWB: the set of detectors used, the local probability of 
the loudest on-source event, and 90\% confidence limits on the gravitational-wave 
amplitude and the distance to the progenitor.  For approximately half of 
the GRBs there was no surviving event and hence no local probability. 
The limits are computed for circularly 
polarized 150 Hz and 1000 Hz sine-Gaussian waveforms.  The distances 
are lower limits, assuming isotropic emission of 
$0.01 M_{\odot}c^2 = 1.8\times10^{53}\mathrm{erg}$ 
of energy in gravitational waves.
These limits include allowances for statistical and systematic errors as 
discussed in Sec.~\ref{sec:errors}.

\begin{deluxetable}{lclrrlrrrrr} 
\tabletypesize{\scriptsize}
\tablecolumns{11} 
\tablewidth{0pc} 
\tablecaption{GRB Sample and Search Results \label{tab:combined}} 
\tablehead{ 
\colhead{}    & \colhead{}    & \colhead{}    & \colhead{}    & 
\colhead{}    & \colhead{}    & \colhead{}    &  \multicolumn{2}{c}{150 Hz} & 
\multicolumn{2}{c}{1000 Hz} \\ 
\cline{8-9} \cline{10-11} \\ 
\colhead{}    & \colhead{}    & \colhead{UTC}    & \colhead{RA}    & 
\colhead{Dec}    & \colhead{}    & \colhead{}    & \colhead{}    & 
\colhead{D}    & \colhead{}    & \colhead{D}    \\ 
\colhead{GRB}    & \colhead{z}   & \colhead{time}   & \colhead{(deg)}   & 
\colhead{(deg)}    & \colhead{network}   & \colhead{$p$}   & \colhead{$h_\mathrm{rss}$}   & 
\colhead{(Mpc)}    & \colhead{$h_\mathrm{rss}$}   & \colhead{(Mpc)} 
}   
\startdata 
051114$\ddagger$ & -- & 04:11:30 & $15^{\mathrm{h}}5^{\mathrm{m}}4^{\mathrm{s}}$ & $60^{\circ}9'$ & H1H2 & -- (162) & 7.98 & 5.7 & 29.9 & 0.229 \\
051117 & -- & 10:51:20 & $15^{\mathrm{h}}13^{\mathrm{m}}36^{\mathrm{s}}$ & $30^{\circ}52'$ & H1H2 & 0.184 (250) & 8.12 & 5.6 & 31.0 & 0.221 \\
051117B & -- & 13:22:54 & $5^{\mathrm{h}}40^{\mathrm{m}}45^{\mathrm{s}}$ & $-19^{\circ}17'$ & H1H2L1 & -- \hphantom{1}(57) & 6.77 & 6.8 & 28.3 & 0.242 \\
051210$\ddagger$ & -- & 05:46:21 & $22^{\mathrm{h}}0^{\mathrm{m}}47^{\mathrm{s}}$ & $-57^{\circ}37'$ & H1H2 & -- (191) & 6.60 & 6.9 & 27.0 & 0.254 \\
051211$\ddagger$ & -- & 02:50:05.4 & $6^{\mathrm{h}}56^{\mathrm{m}}13^{\mathrm{s}}$ & $32^{\circ}41'$ & H1H2L1 & -- (190) & 4.83 & 9.5 & 21.2 & 0.324 \\
051211B & -- & 22:05:44 & $23^{\mathrm{h}}2^{\mathrm{m}}45^{\mathrm{s}}$ & $55^{\circ}5'$ & H1H2L1 & -- (105) & 3.12 & 14.7 & 13.2 & 0.519 \\
051213 & -- & 07:13:04 & $16^{\mathrm{h}}48^{\mathrm{m}}19^{\mathrm{s}}$ & $-59^{\circ}14'$ & H1H2L1 & 0.0769 (104) & 2.62 & 17.5 & 11.3 & 0.609 \\
051221B & -- & 20:03:20 & $20^{\mathrm{h}}49^{\mathrm{m}}26^{\mathrm{s}}$ & $53^{\circ}2'$ & H1H2 & -- \hphantom{1}(82) & 4.91 & 9.3 & 20.6 & 0.334 \\
060102 & -- & 21:17:28 & $21^{\mathrm{h}}55^{\mathrm{m}}20^{\mathrm{s}}$ & $-1^{\circ}50'$ & H1H2 & -- (147) & 6.84 & 6.7 & 27.8 & 0.247 \\
060105 & -- & 06:49:28 & $19^{\mathrm{h}}49^{\mathrm{m}}57^{\mathrm{s}}$ & $46^{\circ}22'$ & H1H2L1 & -- (128) & 5.44 & 8.4 & 23.6 & 0.291 \\
060108 & 2.03 & 14:39:11.76 & $9^{\mathrm{h}}48^{\mathrm{m}}4^{\mathrm{s}}$ & $31^{\circ}56'$ & H1H2L1 & -- \hphantom{1}(89) & 4.92 & 9.3 & 20.4 & 0.336 \\
060110 & -- & 08:01:17 & $4^{\mathrm{h}}50^{\mathrm{m}}57^{\mathrm{s}}$ & $28^{\circ}26'$ & H1H2L1 & -- (135) & 3.58 & 12.8 & 15.5 & 0.444 \\
060111 & -- & 04:23:06 & $18^{\mathrm{h}}24^{\mathrm{m}}47^{\mathrm{s}}$ & $37^{\circ}36'$ & H1H2L1 & -- (131) & 4.97 & 9.2 & 21.1 & 0.325 \\
060114 & -- & 12:39:44 & $13^{\mathrm{h}}1^{\mathrm{m}}7^{\mathrm{s}}$ & $-4^{\circ}45'$ & H1H2L1 & -- (118) & 3.51 & 13.0 & 15.4 & 0.444 \\
060115 & 3.53 & 13:08:00 & $3^{\mathrm{h}}36^{\mathrm{m}}1^{\mathrm{s}}$ & $17^{\circ}20'$ & H1L1 & -- (117) & 4.56 & 10.0 & 20.7 & 0.332 \\
060116 & 6.6 & 08:37:27 & $5^{\mathrm{h}}38^{\mathrm{m}}48^{\mathrm{s}}$ & $-5^{\circ}27'$ & H1H2L1 & 0.0402 (18000) & 5.11 & 9.0 & 26.9 & 0.255 \\
060121$\ddagger$ & -- & 22:24:54.5 & $9^{\mathrm{h}}9^{\mathrm{m}}57^{\mathrm{s}}$ & $45^{\circ}40'$ & H1H2 & -- (159) & 35.32 & 1.3 & 143.6 & 0.048 \\
060202 & -- & 08:40:55 & $2^{\mathrm{h}}23^{\mathrm{m}}17^{\mathrm{s}}$ & $38^{\circ}23'$ & H1H2 & -- (207) & 9.20 & 5.0 & 34.3 & 0.200 \\
060203 & -- & 23:55:35 & $6^{\mathrm{h}}54^{\mathrm{m}}0^{\mathrm{s}}$ & $71^{\circ}50'$ & H1H2 & -- (174) & 6.00 & 7.6 & 21.9 & 0.313 \\
060206 & 4.045 & 04:46:53 & $13^{\mathrm{h}}31^{\mathrm{m}}44^{\mathrm{s}}$ & $35^{\circ}3'$ & H1H2L1 & 0.444 (187) & 4.94 & 9.3 & 21.9 & 0.313 \\
060211B & -- & 15:55:15 & $5^{\mathrm{h}}0^{\mathrm{m}}18^{\mathrm{s}}$ & $14^{\circ}57'$ & H1H2 & -- (149) & 8.67 & 5.3 & 29.0 & 0.237 \\
060223 & 4.41 & 06:04:23 & $3^{\mathrm{h}}40^{\mathrm{m}}45^{\mathrm{s}}$ & $-17^{\circ}8'$ & H1H2L1 & 0.321 (162) & 4.88 & 9.4 & 21.0 & 0.327 \\
060306 & -- & 00:49:10 & $2^{\mathrm{h}}44^{\mathrm{m}}23^{\mathrm{s}}$ & $-2^{\circ}9'$ & H1H2L1 & 0.102 (186) & 3.45 & 13.3 & 15.1 & 0.454 \\
060312 & -- & 01:36:12 & $3^{\mathrm{h}}3^{\mathrm{m}}6^{\mathrm{s}}$ & $12^{\circ}49'$ & H1H2L1 & -- (196) & 3.14 & 14.6 & 11.8 & 0.581 \\
060313$\ddagger$ & $<$1.7 & 00:12:06 & $4^{\mathrm{h}}26^{\mathrm{m}}30^{\mathrm{s}}$ & $-10^{\circ}52'$ & H1H2 & -- (186) & 4.92 & 9.3 & 20.5 & 0.335 \\
060319 & -- & 00:55:42 & $11^{\mathrm{h}}45^{\mathrm{m}}28^{\mathrm{s}}$ & $60^{\circ}2'$ & H1H2 & -- (187) & 4.90 & 9.3 & 20.8 & 0.331 \\
060323 & -- & 14:32:36 & $11^{\mathrm{h}}37^{\mathrm{m}}39^{\mathrm{s}}$ & $50^{\circ}0'$ & H1H2 & -- \hphantom{1}(84) & 5.26 & 8.7 & 22.1 & 0.311 \\
060403 & -- & 13:12:17 & $18^{\mathrm{h}}49^{\mathrm{m}}21^{\mathrm{s}}$ & $8^{\circ}20'$ & H1H2 & -- (207) & 3.66 & 12.5 & 15.3 & 0.450 \\
060418 & 1.49 & 03:06:08 & $15^{\mathrm{h}}45^{\mathrm{m}}43^{\mathrm{s}}$ & $-3^{\circ}39'$ & H1H2L1 & 0.681 (141) & 7.01 & 6.5 & 34.1 & 0.201 \\
060427 & -- & 11:43:10 & $8^{\mathrm{h}}16^{\mathrm{m}}42^{\mathrm{s}}$ & $62^{\circ}39'$ & H1H2L1 & -- (168) & 4.60 & 9.9 & 20.5 & 0.334 \\
060427B$\ddagger$ & -- & 23:51:55 & $6^{\mathrm{h}}33^{\mathrm{m}}53^{\mathrm{s}}$ & $21^{\circ}21'$ & H1H2L1 & 0.228 (114) & 2.44 & 18.7 & 10.6 & 0.649 \\
060428 & -- & 03:22:48 & $8^{\mathrm{h}}14^{\mathrm{m}}8^{\mathrm{s}}$ & $-37^{\circ}10'$ & H1H2 & -- (207) & 18.52 & 2.5 & 79.7 & 0.086 \\
060428B & -- & 08:54:38 & $15^{\mathrm{h}}41^{\mathrm{m}}31^{\mathrm{s}}$ & $62^{\circ}2'$ & H1H2L1 & 0.0139 (18000) & 2.39 & 19.2 & 10.8 & 0.637 \\
060429$\ddagger$ & -- & 12:19:51.00 & $7^{\mathrm{h}}42^{\mathrm{m}}3^{\mathrm{s}}$ & $-24^{\circ}57'$ & H1H2 & -- (166) & 3.36 & 13.6 & 14.5 & 0.472 \\
060501 & -- & 08:14:58 & $21^{\mathrm{h}}53^{\mathrm{m}}32^{\mathrm{s}}$ & $43^{\circ}60'$ & H1H2 & -- (172) & 5.72 & 8.0 & 24.0 & 0.286 \\
060510 & -- & 07:43:27 & $6^{\mathrm{h}}23^{\mathrm{m}}25^{\mathrm{s}}$ & $-1^{\circ}10'$ & H1H2L1 & -- (114) & 3.36 & 13.6 & 15.1 & 0.455 \\
060510B & 4.9 & 08:22:14 & $15^{\mathrm{h}}56^{\mathrm{m}}52^{\mathrm{s}}$ & $78^{\circ}36'$ & H1H2L1 & 0.0124 (18000) & 2.89 & 15.9 & 14.7 & 0.466 \\
060515 & -- & 02:27:52 & $8^{\mathrm{h}}29^{\mathrm{m}}11^{\mathrm{s}}$ & $73^{\circ}34'$ & H1H2L1 & 0.509 \hphantom{1}(57) & 2.36 & 19.4 & 10.6 & 0.650 \\
060516 & -- & 06:43:34 & $4^{\mathrm{h}}44^{\mathrm{m}}40^{\mathrm{s}}$ & $-18^{\circ}6'$ & H1H2L1 & 0.221 (140) & 2.09 & 21.9 & 10.4 & 0.657 \\
060526 & 3.21 & 16:28:30 & $15^{\mathrm{h}}31^{\mathrm{m}}21^{\mathrm{s}}$ & $0^{\circ}18'$ & H1H2L1 & 0.731 \hphantom{1}(52) & 2.56 & 17.9 & 11.7 & 0.587 \\
060605 & 3.8 & 18:15:44 & $21^{\mathrm{h}}28^{\mathrm{m}}38^{\mathrm{s}}$ & $-6^{\circ}3'$ & H1H2 & -- (201) & 10.63 & 4.3 & 43.3 & 0.158 \\
060607 & -- & 05:12:13 & $21^{\mathrm{h}}58^{\mathrm{m}}51^{\mathrm{s}}$ & $-22^{\circ}30'$ & H1H2L1 & 0.0945 (201) & 4.88 & 9.4 & 21.9 & 0.314 \\
060607B & -- & 23:32:44 & $2^{\mathrm{h}}48^{\mathrm{m}}12^{\mathrm{s}}$ & $14^{\circ}45'$ & H1H2L1 & -- (135) & 9.49 & 4.8 & 41.2 & 0.167 \\
060614 & 0.125 & 12:43:48 & $21^{\mathrm{h}}23^{\mathrm{m}}31^{\mathrm{s}}$ & $-53^{\circ}2'$ & H2L1 & -- \hphantom{1}(61) & 26.59 & 1.7 & 118.8 & 0.058 \\
060707 & 3.43 & 21:30:19 & $23^{\mathrm{h}}48^{\mathrm{m}}18^{\mathrm{s}}$ & $-17^{\circ}54'$ & H1H2L1 & -- (188) & 2.53 & 18.1 & 11.2 & 0.612 \\
060712 & -- & 21:07:43 & $12^{\mathrm{h}}16^{\mathrm{m}}16^{\mathrm{s}}$ & $35^{\circ}32'$ & H1H2 & -- \hphantom{1}(65) & 4.89 & 9.4 & 19.9 & 0.344 \\
060714 & 2.71 & 15:12:00 & $15^{\mathrm{h}}11^{\mathrm{m}}25^{\mathrm{s}}$ & $-6^{\circ}33'$ & H1H2 & -- (162) & 3.88 & 11.8 & 15.6 & 0.440 \\
060719 & -- & 06:50:36 & $1^{\mathrm{h}}13^{\mathrm{m}}40^{\mathrm{s}}$ & $-48^{\circ}23'$ & H1H2L1 & -- (127) & 3.46 & 13.2 & 15.3 & 0.447 \\
060804 & -- & 07:28:19 & $7^{\mathrm{h}}28^{\mathrm{m}}52^{\mathrm{s}}$ & $-27^{\circ}14'$ & H1H2L1 & -- (109) & 2.34 & 19.5 & 10.8 & 0.635 \\
060805 & -- & 04:47:49 & $14^{\mathrm{h}}43^{\mathrm{m}}42^{\mathrm{s}}$ & $12^{\circ}35'$ & H1H2L1 & 0.569 (195) & 3.34 & 13.7 & 14.9 & 0.462 \\
060807 & -- & 14:41:35 & $16^{\mathrm{h}}50^{\mathrm{m}}1^{\mathrm{s}}$ & $31^{\circ}36'$ & H1H2L1 & 0.00967 (18000) & 4.70 & 9.7 & 21.2 & 0.323 \\
060813 & -- & 22:50:22 & $7^{\mathrm{h}}27^{\mathrm{m}}34^{\mathrm{s}}$ & $-29^{\circ}51'$ & H1H2L1 & 0.297 (185) & 3.49 & 13.1 & 15.8 & 0.434 \\
060814 & 0.84 & 23:02:19 & $14^{\mathrm{h}}45^{\mathrm{m}}21^{\mathrm{s}}$ & $20^{\circ}36'$ & H2L1 & 0.0741 (135) & 3.21 & 14.2 & 13.5 & 0.507 \\
060825 & -- & 02:59:57 & $1^{\mathrm{h}}12^{\mathrm{m}}31^{\mathrm{s}}$ & $55^{\circ}48'$ & H1H2 & -- (163) & 5.06 & 9.1 & 21.9 & 0.313 \\
060904 & -- & 01:03:21 & $15^{\mathrm{h}}50^{\mathrm{m}}55^{\mathrm{s}}$ & $44^{\circ}59'$ & H1H2L1 & -- (146) & 1.82 & 25.1 & 8.1 & 0.843 \\
060904B & 0.703 & 02:31:03 & $3^{\mathrm{h}}52^{\mathrm{m}}52^{\mathrm{s}}$ & $0^{\circ}44'$ & H1H2 & 0.391 (179) & 3.57 & 12.8 & 15.3 & 0.447 \\
060906 & 3.685 & 08:32:46 & $2^{\mathrm{h}}42^{\mathrm{m}}50^{\mathrm{s}}$ & $30^{\circ}21'$ & H1H2L1 & -- (187) & 2.30 & 19.9 & 10.6 & 0.647 \\
060908 & 2.43 & 08:57:22 & $2^{\mathrm{h}}7^{\mathrm{m}}17^{\mathrm{s}}$ & $0^{\circ}20'$ & H1H2L1 & 0.487 (189) & 2.34 & 19.6 & 10.9 & 0.631 \\
060919 & -- & 07:48:38 & $18^{\mathrm{h}}27^{\mathrm{m}}36^{\mathrm{s}}$ & $-50^{\circ}60'$ & H1H2L1 & 0.130 (138) & 3.26 & 14.0 & 15.1 & 0.456 \\
060923 & -- & 05:12:15 & $16^{\mathrm{h}}58^{\mathrm{m}}30^{\mathrm{s}}$ & $12^{\circ}20'$ & H1H2 & -- (142) & 5.11 & 9.0 & 22.3 & 0.308 \\
060923C & -- & 13:33:02 & $23^{\mathrm{h}}4^{\mathrm{m}}29^{\mathrm{s}}$ & $3^{\circ}57'$ & H1H2 & -- (199) & 37.92 & 1.2 & 164.5 & 0.042 \\
060927 & 5.6 & 14:07:35 & $21^{\mathrm{h}}58^{\mathrm{m}}11^{\mathrm{s}}$ & $5^{\circ}22'$ & H1H2 & 0.576 (184) & 4.79 & 9.6 & 21.1 & 0.325 \\
060928 & -- & 01:17:01.00 & $8^{\mathrm{h}}30^{\mathrm{m}}27^{\mathrm{s}}$ & $-42^{\circ}44'$ & H1H2 & 0.228 (114) & 2.96 & 15.5 & 11.7 & 0.587 \\
\enddata 
\end{deluxetable}

\begin{deluxetable}{lclrrlrrrrr} 
\tabletypesize{\scriptsize}
\tablecolumns{11} 
\tablewidth{0pc} 
\tablenum{1}
\tablecaption{$-$ {\em Continued}}
\tablehead{ 
\colhead{}    & \colhead{}    & \colhead{}    & \colhead{}    & 
\colhead{}    & \colhead{}    & \colhead{}    &  \multicolumn{2}{c}{150 Hz} & 
\multicolumn{2}{c}{1000 Hz} \\ 
\cline{8-9} \cline{10-11} \\ 
\colhead{}    & \colhead{}    & \colhead{UTC}    & \colhead{RA}    & 
\colhead{Dec}    & \colhead{}    & \colhead{}    & \colhead{}    & 
\colhead{D}    & \colhead{}    & \colhead{D}    \\ 
\colhead{GRB}    & \colhead{z}   & \colhead{time}   & \colhead{(deg)}   & 
\colhead{(deg)}    & \colhead{network}   & \colhead{$p$}   & \colhead{$h_\mathrm{rss}$}   & 
\colhead{(Mpc)}    & \colhead{$h_\mathrm{rss}$}   & \colhead{(Mpc)} 
}   
\startdata 
060930 & -- & 09:04:09 & $20^{\mathrm{h}}18^{\mathrm{m}}9^{\mathrm{s}}$ & $-23^{\circ}38'$ & H1L1 & 0.0248 (18000) & 6.95 & 6.6 & 36.9 & 0.186 \\
061002 & -- & 01:03:29 & $14^{\mathrm{h}}41^{\mathrm{m}}25^{\mathrm{s}}$ & $48^{\circ}44'$ & H1H2L1 & -- (193) & 2.49 & 18.3 & 11.2 & 0.615 \\
061006$\ddagger$ & -- & 16:45:50 & $7^{\mathrm{h}}23^{\mathrm{m}}60^{\mathrm{s}}$ & $-79^{\circ}12'$ & H1H2 & 0.310 (184) & 3.61 & 12.7 & 18.8 & 0.365 \\
061007 & 1.261 & 10:08:08 & $3^{\mathrm{h}}5^{\mathrm{m}}12^{\mathrm{s}}$ & $-50^{\circ}30'$ & H1H2L1 & 0.775 (160) & 9.70 & 4.7 & 42.7 & 0.161 \\
061021 & $<$2.0 & 15:39:07 & $9^{\mathrm{h}}40^{\mathrm{m}}35^{\mathrm{s}}$ & $-21^{\circ}57'$ & H1H2L1 & 0.979 \hphantom{1}(94) & 4.32 & 10.6 & 19.8 & 0.347 \\
061027 & -- & 10:15:02 & $18^{\mathrm{h}}3^{\mathrm{m}}58^{\mathrm{s}}$ & $-82^{\circ}14'$ & H1H2 & -- (193) & 4.42 & 10.4 & 15.4 & 0.446 \\
061102 & -- & 01:00:31 & $9^{\mathrm{h}}53^{\mathrm{m}}34^{\mathrm{s}}$ & $-17^{\circ}0'$ & H1H2L1 & -- (113) & 2.38 & 19.2 & 10.7 & 0.639 \\
061110 & 0.757 & 11:47:21 & $22^{\mathrm{h}}25^{\mathrm{m}}8^{\mathrm{s}}$ & $-2^{\circ}15'$ & H1H2L1 & 0.214 (168) & 3.12 & 14.6 & 14.1 & 0.486 \\
061122 & -- & 07:56:49 & $20^{\mathrm{h}}15^{\mathrm{m}}21^{\mathrm{s}}$ & $15^{\circ}31'$ & H1H2L1 & 0.575 \hphantom{1}(73) & 4.36 & 10.5 & 20.6 & 0.334 \\
061126 & $<$1.5 & 08:47:56 & $5^{\mathrm{h}}46^{\mathrm{m}}28^{\mathrm{s}}$ & $64^{\circ}12'$ & H1H2 & -- (144) & 2.79 & 16.4 & 11.0 & 0.622 \\
061201$\ddagger$ & -- & 15:58:36 & $22^{\mathrm{h}}8^{\mathrm{m}}19^{\mathrm{s}}$ & $-74^{\circ}34'$ & H1H2 & 0.0222 (18000) & 3.53 & 13.0 & 16.8 & 0.408 \\
061217$\ddagger$ & 0.827 & 03:40:08 & $10^{\mathrm{h}}41^{\mathrm{m}}40^{\mathrm{s}}$ & $-21^{\circ}9'$ & H1L1 & -- (187) & 3.32 & 13.8 & 15.8 & 0.433 \\
061218 & -- & 04:05:05 & $9^{\mathrm{h}}56^{\mathrm{m}}57^{\mathrm{s}}$ & $-35^{\circ}13'$ & H1H2L1 & -- (169) & 3.67 & 12.5 & 15.9 & 0.431 \\
061222 & -- & 03:28:52 & $23^{\mathrm{h}}53^{\mathrm{m}}2^{\mathrm{s}}$ & $46^{\circ}32'$ & H1H2 & -- (207) & 4.85 & 9.4 & 15.9 & 0.430 \\
061222B & 3.355 & 04:11:02 & $7^{\mathrm{h}}1^{\mathrm{m}}24^{\mathrm{s}}$ & $-25^{\circ}52'$ & H1H2L1 & 0.444 (180) & 6.50 & 7.0 & 28.0 & 0.245 \\
070103 & -- & 20:46:39.41 & $23^{\mathrm{h}}30^{\mathrm{m}}20^{\mathrm{s}}$ & $26^{\circ}49'$ & H1H2 & -- (207) & 5.37 & 8.5 & 21.4 & 0.321 \\
070107 & -- & 12:05:18 & $10^{\mathrm{h}}37^{\mathrm{m}}41^{\mathrm{s}}$ & $-53^{\circ}12'$ & H1H2L1 & -- (186) & 15.57 & 2.9 & 60.2 & 0.114 \\
070110 & 2.352 & 07:22:41 & $0^{\mathrm{h}}3^{\mathrm{m}}44^{\mathrm{s}}$ & $-52^{\circ}59'$ & H1H2L1 & 0.609 (207) & 2.50 & 18.3 & 11.1 & 0.618 \\
070129 & -- & 23:35:10 & $2^{\mathrm{h}}28^{\mathrm{m}}0^{\mathrm{s}}$ & $11^{\circ}44'$ & H1H2 & 0.261 (207) & 3.50 & 13.1 & 14.8 & 0.462 \\
070201$\ddagger$ & -- & 15:23:10.78 & $0^{\mathrm{h}}44^{\mathrm{m}}21^{\mathrm{s}}$ & $42^{\circ}18'$ & H1H2 & 0.0791 (177) & 6.38 & 7.2 & 27.8 & 0.247 \\
070208 & 1.165 & 09:10:34 & $13^{\mathrm{h}}11^{\mathrm{m}}35^{\mathrm{s}}$ & $61^{\circ}57'$ & H1H2L1 & 0.0847 (177) & 1.87 & 24.5 & 10.4 & 0.658 \\
070209$\ddagger$ & -- & 03:33:41 & $3^{\mathrm{h}}4^{\mathrm{m}}51^{\mathrm{s}}$ & $-47^{\circ}23'$ & H1H2L1 & 0.605 (185) & 11.24 & 4.1 & 52.6 & 0.131 \\
070219 & -- & 01:10:16 & $17^{\mathrm{h}}20^{\mathrm{m}}53^{\mathrm{s}}$ & $69^{\circ}21'$ & H1L1 & 0.192 (104) & 3.61 & 12.7 & 20.6 & 0.334 \\
070223 & -- & 01:15:00 & $10^{\mathrm{h}}13^{\mathrm{m}}49^{\mathrm{s}}$ & $43^{\circ}8'$ & H1H2L1 & 0.219 (137) & 3.36 & 13.6 & 15.3 & 0.448 \\
070309 & -- & 10:01:03 & $17^{\mathrm{h}}34^{\mathrm{m}}44^{\mathrm{s}}$ & $-37^{\circ}57'$ & H1H2L1 & 0.357 (196) & 3.92 & 11.7 & 18.6 & 0.370 \\
070311 & -- & 01:52:35 & $5^{\mathrm{h}}50^{\mathrm{m}}10^{\mathrm{s}}$ & $3^{\circ}23'$ & H1H2L1 & 0.447 (188) & 2.35 & 19.5 & 10.9 & 0.631 \\
070318 & 0.836 & 07:28:56 & $3^{\mathrm{h}}13^{\mathrm{m}}57^{\mathrm{s}}$ & $-42^{\circ}57'$ & H1H2L1 & 0.873 (166) & 2.14 & 21.4 & 10.1 & 0.680 \\
070330 & -- & 22:51:31 & $17^{\mathrm{h}}58^{\mathrm{m}}8^{\mathrm{s}}$ & $-63^{\circ}48'$ & H1H2L1 & 0.134 (201) & 1.87 & 24.5 & 10.2 & 0.671 \\
070402 & -- & 15:48:35.00 & $20^{\mathrm{h}}44^{\mathrm{m}}44^{\mathrm{s}}$ & $27^{\circ}24'$ & H1H2L1 & 0.299 \hphantom{1}(87) & 2.24 & 20.5 & 10.3 & 0.667 \\
070411 & 2.954 & 20:12:33 & $7^{\mathrm{h}}9^{\mathrm{m}}23^{\mathrm{s}}$ & $1^{\circ}3'$ & H2L1 & 0.0733 (150) & 18.35 & 2.5 & 75.3 & 0.091 \\
070412 & -- & 01:27:03 & $12^{\mathrm{h}}6^{\mathrm{m}}6^{\mathrm{s}}$ & $40^{\circ}8'$ & H1H2L1 & 0.915 (177) & 2.49 & 18.4 & 11.1 & 0.621 \\
070419 & 0.97 & 09:59:26 & $12^{\mathrm{h}}11^{\mathrm{m}}1^{\mathrm{s}}$ & $39^{\circ}54'$ & H1H2L1 & 0.715 (123) & 2.75 & 16.6 & 12.8 & 0.535 \\
070419B & -- & 10:44:05 & $21^{\mathrm{h}}2^{\mathrm{m}}50^{\mathrm{s}}$ & $-31^{\circ}16'$ & H1H2 & -- (193) & 6.14 & 7.4 & 24.9 & 0.276 \\
070420 & -- & 06:18:13 & $8^{\mathrm{h}}4^{\mathrm{m}}59^{\mathrm{s}}$ & $-45^{\circ}34'$ & H1H2L1 & 0.805 (133) & 3.58 & 12.8 & 18.8 & 0.365 \\
070427 & -- & 08:31:08 & $1^{\mathrm{h}}55^{\mathrm{m}}29^{\mathrm{s}}$ & $-27^{\circ}36'$ & H1H2L1 & -- (150) & 2.02 & 22.6 & 10.1 & 0.679 \\
070429 & -- & 01:35:10 & $19^{\mathrm{h}}50^{\mathrm{m}}47^{\mathrm{s}}$ & $-32^{\circ}25'$ & H1L1 & -- (152) & 1.79 & 25.6 & 10.6 & 0.647 \\
070429B$\ddagger$ & -- & 03:09:04 & $21^{\mathrm{h}}52^{\mathrm{m}}1^{\mathrm{s}}$ & $-38^{\circ}51'$ & H1H2L1 & 0.443 (194) & 1.75 & 26.2 & 8.0 & 0.862 \\
070506 & 2.31 & 05:35:58 & $23^{\mathrm{h}}8^{\mathrm{m}}49^{\mathrm{s}}$ & $10^{\circ}43'$ & H1H2L1 & 0.811 (122) & 3.17 & 14.4 & 15.3 & 0.450 \\
070508 & $<$2.3 & 04:18:17 & $20^{\mathrm{h}}51^{\mathrm{m}}20^{\mathrm{s}}$ & $-78^{\circ}23'$ & H1H2L1 & 0.147 (184) & 2.37 & 19.3 & 10.7 & 0.642 \\
070518 & -- & 14:26:21 & $16^{\mathrm{h}}56^{\mathrm{m}}53^{\mathrm{s}}$ & $55^{\circ}17'$ & H1H2 & 0.525 (120) & 3.48 & 13.1 & 15.1 & 0.453 \\
070520 & -- & 13:05:10 & $12^{\mathrm{h}}53^{\mathrm{m}}1^{\mathrm{s}}$ & $75^{\circ}0'$ & H1H2V1 & -- (180) & 4.20 & 10.9 & 16.2 & 0.424 \\
070520B & -- & 17:44:53 & $8^{\mathrm{h}}7^{\mathrm{m}}33^{\mathrm{s}}$ & $57^{\circ}35'$ & L1V1 & 0.487 (195) & 22.37 & 2.0 & 30.0 & 0.229 \\
070521 & -- & 06:51:10 & $16^{\mathrm{h}}10^{\mathrm{m}}38^{\mathrm{s}}$ & $30^{\circ}16'$ & H1H2V1 & -- (167) & 2.62 & 17.5 & 12.1 & 0.569 \\
070529 & 2.4996 & 12:48:28 & $18^{\mathrm{h}}54^{\mathrm{m}}54^{\mathrm{s}}$ & $20^{\circ}39'$ & H1H2L1V1 & 0.0776 (18000) & 2.86 & 16.0 & 13.0 & 0.528 \\
070531 & -- & 02:10:17 & $0^{\mathrm{h}}26^{\mathrm{m}}53^{\mathrm{s}}$ & $74^{\circ}19'$ & L1V1 & 0.533 (184) & 11.47 & 4.0 & 18.4 & 0.372 \\
070611 & 2.04 & 01:57:13 & $0^{\mathrm{h}}8^{\mathrm{m}}1^{\mathrm{s}}$ & $-29^{\circ}45'$ & H1H2L1 & -- (172) & 1.80 & 25.4 & 8.5 & 0.805 \\
070612 & 0.617 & 02:38:45 & $8^{\mathrm{h}}5^{\mathrm{m}}25^{\mathrm{s}}$ & $37^{\circ}15'$ & L1V1 & 0.174 (207) & 13.85 & 3.3 & 68.3 & 0.100 \\
070612B & -- & 06:21:17 & $17^{\mathrm{h}}26^{\mathrm{m}}52^{\mathrm{s}}$ & $-8^{\circ}45'$ & H1H2L1V1 & 0.129 (124) & 2.62 & 17.5 & 14.4 & 0.477 \\
070615 & -- & 02:20:35 & $2^{\mathrm{h}}57^{\mathrm{m}}14^{\mathrm{s}}$ & $-4^{\circ}24'$ & H1H2L1V1 & 0.219 (169) & 2.57 & 17.8 & 11.2 & 0.614 \\
070616 & -- & 16:29:33 & $2^{\mathrm{h}}8^{\mathrm{m}}23^{\mathrm{s}}$ & $56^{\circ}57'$ & H1H2V1 & 0.633 (166) & 2.79 & 16.4 & 10.8 & 0.636 \\
070621 & -- & 23:17:39 & $21^{\mathrm{h}}35^{\mathrm{m}}13^{\mathrm{s}}$ & $-24^{\circ}49'$ & H1L1V1 & 0.652 \hphantom{1}(69) & 1.79 & 25.6 & 10.0 & 0.689 \\
070626 & -- & 04:05:33 & $9^{\mathrm{h}}25^{\mathrm{m}}25^{\mathrm{s}}$ & $-39^{\circ}52'$ & H1L1V1 & -- \hphantom{1}(86) & 4.96 & 9.2 & 18.6 & 0.368 \\
070628 & -- & 14:41:02 & $7^{\mathrm{h}}41^{\mathrm{m}}5^{\mathrm{s}}$ & $-20^{\circ}17'$ & H1V1 & 0.767 (133) & 9.88 & 4.6 & 43.7 & 0.157 \\
070704 & -- & 20:05:57 & $23^{\mathrm{h}}38^{\mathrm{m}}50^{\mathrm{s}}$ & $66^{\circ}15'$ & H1H2L1 & 0.237 \hphantom{1}(80) & 3.66 & 12.5 & 15.1 & 0.455 \\
070707$\ddagger$ & -- & 16:08:38 & $17^{\mathrm{h}}51^{\mathrm{m}}0^{\mathrm{s}}$ & $-68^{\circ}53'$ & L1V1 & 0.799 (184) & 36.04 & 1.3 & 85.5 & 0.080 \\
070714$\ddagger$ & -- & 03:20:31 & $2^{\mathrm{h}}51^{\mathrm{m}}44^{\mathrm{s}}$ & $30^{\circ}14'$ & H1H2L1V1 & -- (114) & 5.04 & 9.1 & 26.0 & 0.264 \\
070714B$\ddagger$ & 0.92 & 04:59:29 & $3^{\mathrm{h}}51^{\mathrm{m}}25^{\mathrm{s}}$ & $28^{\circ}18'$ & H1H2L1V1 & 0.965 (141) & 5.46 & 8.4 & 20.7 & 0.331 \\
070721 & -- & 10:01:08 & $0^{\mathrm{h}}12^{\mathrm{m}}35^{\mathrm{s}}$ & $-28^{\circ}32'$ & H1H2L1V1 & -- (138) & 4.29 & 10.7 & 15.2 & 0.450 \\
070721B & 3.626 & 10:33:48 & $2^{\mathrm{h}}12^{\mathrm{m}}31^{\mathrm{s}}$ & $-2^{\circ}12'$ & H1H2L1V1 & 0.492 (118) & 3.49 & 13.1 & 15.3 & 0.450 \\
070724$\ddagger$ & 0.457 & 10:53:50 & $1^{\mathrm{h}}51^{\mathrm{m}}18^{\mathrm{s}}$ & $-18^{\circ}37'$ & H1H2L1V1 & 0.191 (110) & 4.76 & 9.6 & 19.2 & 0.357 \\
070724B & -- & 23:25:09 & $1^{\mathrm{h}}10^{\mathrm{m}}31^{\mathrm{s}}$ & $57^{\circ}40'$ & H1H2L1V1 & -- (164) & 4.85 & 9.4 & 19.9 & 0.344 \\
070729$\ddagger$ & -- & 00:25:53 & $3^{\mathrm{h}}45^{\mathrm{m}}11^{\mathrm{s}}$ & $-39^{\circ}20'$ & H1H2L1V1 & -- (155) & 2.33 & 19.6 & 10.7 & 0.639 \\
070731 & -- & 09:33:22 & $21^{\mathrm{h}}54^{\mathrm{m}}19^{\mathrm{s}}$ & $-15^{\circ}44'$ & H1H2L1V1 & -- \hphantom{1}(84) & 4.97 & 9.2 & 16.7 & 0.410 \\
\enddata 
\end{deluxetable}

\begin{deluxetable}{lclrrlrrrrr} 
\tabletypesize{\scriptsize}
\tablecolumns{11} 
\tablewidth{0pc} 
\tablenum{1}
\tablecaption{$-$ {\em Continued}}
\tablehead{ 
\colhead{}    & \colhead{}    & \colhead{}    & \colhead{}    & 
\colhead{}    & \colhead{}    & \colhead{}    &  \multicolumn{2}{c}{150 Hz} & 
\multicolumn{2}{c}{1000 Hz} \\ 
\cline{8-9} \cline{10-11} \\ 
\colhead{}    & \colhead{}    & \colhead{UTC}    & \colhead{RA}    & 
\colhead{Dec}    & \colhead{}    & \colhead{}    & \colhead{}    & 
\colhead{D}    & \colhead{}    & \colhead{D}    \\ 
\colhead{GRB}    & \colhead{z}   & \colhead{time}   & \colhead{(deg)}   & 
\colhead{(deg)}    & \colhead{network}   & \colhead{$p$}   & \colhead{$h_\mathrm{rss}$}   & 
\colhead{(Mpc)}    & \colhead{$h_\mathrm{rss}$}   & \colhead{(Mpc)} 
}   
\startdata 
070802 & 2.45 & 07:07:25 & $2^{\mathrm{h}}27^{\mathrm{m}}37^{\mathrm{s}}$ & $-55^{\circ}31'$ & H1H2 & -- (161) & 3.56 & 12.8 & 15.4 & 0.445 \\
070805 & -- & 19:55:45 & $16^{\mathrm{h}}20^{\mathrm{m}}14^{\mathrm{s}}$ & $-59^{\circ}57'$ & H1H2L1 & 0.193 (207) & 2.51 & 18.2 & 13.9 & 0.493 \\
070809$\ddagger$ & -- & 19:22:17 & $13^{\mathrm{h}}35^{\mathrm{m}}4^{\mathrm{s}}$ & $-22^{\circ}7'$ & H1H2V1 & -- (183) & 9.20 & 5.0 & 38.6 & 0.178 \\
070810 & 2.17 & 02:11:52 & $12^{\mathrm{h}}39^{\mathrm{m}}47^{\mathrm{s}}$ & $10^{\circ}45'$ & H1H2L1V1 & -- (120) & 4.48 & 10.2 & 15.4 & 0.446 \\
070810B$\ddagger$ & -- & 15:19:17 & $0^{\mathrm{h}}35^{\mathrm{m}}48^{\mathrm{s}}$ & $8^{\circ}49'$ & H1H2L1 & 0.239 (180) & 4.80 & 9.5 & 21.4 & 0.321 \\
070821 & -- & 12:49:24.00 & $6^{\mathrm{h}}22^{\mathrm{m}}6^{\mathrm{s}}$ & $-63^{\circ}51'$ & H1H2L1V1 & 0.303 (119) & 3.96 & 11.6 & 14.7 & 0.467 \\
070911 & -- & 05:57:44 & $1^{\mathrm{h}}43^{\mathrm{m}}17^{\mathrm{s}}$ & $-33^{\circ}29'$ & H1H2L1V1 & 0.512 (160) & 4.74 & 9.6 & 17.7 & 0.387 \\
070917 & -- & 07:33:56 & $19^{\mathrm{h}}35^{\mathrm{m}}42^{\mathrm{s}}$ & $2^{\circ}25'$ & H1H2V1 & 0.295 (193) & 4.23 & 10.8 & 15.6 & 0.439 \\
070920 & -- & 04:00:13 & $6^{\mathrm{h}}43^{\mathrm{m}}52^{\mathrm{s}}$ & $72^{\circ}15'$ & H1H2L1V1 & -- (123) & 3.54 & 12.9 & 15.6 & 0.441 \\
070920B & -- & 21:04:32 & $0^{\mathrm{h}}0^{\mathrm{m}}30^{\mathrm{s}}$ & $-34^{\circ}51'$ & H1H2L1V1 & 0.600 \hphantom{1}(60) & 2.26 & 20.2 & 10.4 & 0.659 \\
070923$\ddagger$ & -- & 19:15:23 & $12^{\mathrm{h}}18^{\mathrm{m}}30^{\mathrm{s}}$ & $-38^{\circ}18'$ & H1H2L1 & -- (196) & 4.91 & 9.3 & 26.0 & 0.264 \\
\enddata 

\tablecomments{
Information and limits on associated GWB emission for each of the 
GRBs studied. The first five columns are GRB name, redshift (if known), time, 
and sky position (right ascension and declination).  The remaining columns 
display the results of the \xp~search for an associated GWB: the set of 
detectors used, the local probability $p$ of the loudest on-source event, 
and 90\% confidence limits on the gravitational-wave amplitude and the distance to the progenitor. 
A $p$ value of ``$-$'' indicates no event survived all cuts.  
The number in parentheses after the $p$ value is the 
number of off-source segments used to estimate $p$.  
The limits are computed for circularly 
polarized 150 Hz and 1000 Hz sine-Gaussian waveforms.  
The $h_\mathrm{rss}$ amplitudes are in units of $10^{-22}\mathrm{Hz}^{-1/2}$.
The distances are lower limits, assuming isotropic emission of 
$E_{\mathrm{GW}}^{\mathrm{iso}}=0.01 M_{\odot}c^2 = 1.8\times10^{52}\mathrm{erg}$ 
in gravitational waves, and scale as 
$D\propto(E_{\mathrm{GW}}^{\mathrm{iso}})^{1/2}$. 
These limits include allowances for systematics as discussed in Sec.~\ref{sec:errors}. 
A double dagger ($\ddagger$) following the GRB name indicates that it was also 
included in the template-based search for binary inspiral gravitational-wave 
signals presented in \citet{CBCGRB}.
}
\end{deluxetable}

\end{document}